\documentclass[11pt]{article}
\usepackage{amssymb}
\usepackage{amsmath}
\usepackage{amsfonts}
\usepackage[left=2.6cm,right=2.6cm,top=2.0cm,bottom=2.0cm]{geometry}
\usepackage[left=2.6cm,right=2.6cm,top=2.0cm,bottom=2.0cm]{geometry}
\usepackage{lscape}
\usepackage{utopia}
\usepackage[doublespacing]{setspace}
\usepackage{boxedminipage}
\usepackage{epsfig}
\usepackage{float}
\usepackage{caption}
\usepackage{hyperref}

\graphicspath{{./Graphics/}}

\begin{document}

\title{A New Test for Market Efficiency\\ and Uncovered Interest Parity}
\author{Richard T. Baillie \\
%EndAName
Michigan State University, USA\\
King's College, University of London, UK \bigskip \\ 
Francis X. Diebold \\
University of Pennsylvania, USA  \bigskip \\
George Kapetanios \\
King's College, University of London, UK  \bigskip \\
Kun Ho Kim \\
Concordia University, Canada 
}

\date{}
\maketitle

 \thispagestyle{empty} 

\vspace{-1cm}

\begin{singlespace}

\begin{center}

First Draft: September 26, 2021

\smallskip

This Draft: \today

\end{center}
	
	\bigskip
	
\noindent {Abstract}:  We suggest a new single-equation test for Uncovered Interest Parity
($UIP$) based on a dynamic regression approach. The method provides
consistent and asymptotically efficient parameter estimates, and is not
dependent on assumptions of strict exogeneity. This new approach is
asymptotically more efficient than the common approach of using $OLS$ with $%
HAC$ robust standard errors in the static forward premium regression. The
coefficient estimates when spot return changes are regressed on the forward premium
 are all positive and remarkably stable across currencies. These estimates are considerably larger than those of 
previous studies, which frequently find negative coefficients. The method also has the 
 advantage of showing dynamic effects of risk premia, or other
events that may lead to rejection of $UIP$ or the efficient markets
hypothesis.

\bigskip


\bigskip


\bigskip

\noindent Acknowledgments: For helpful comments we thank the co-editor and two referees.  The usual disclaimer applies.

\end{singlespace}

\clearpage

\setcounter{page}{1}
\thispagestyle{empty}

\section{Introduction}

Considerable past research in international finance tests the major parity
conditions and/or models the role of risk premia and informational
inefficiency in currency markets. This paper introduces a new and simple
single-equation approach for testing uncovered interest parity ($UIP$),
which also allows for the inclusion of other variables that could represent
time-varying risk premia. Our approach is based on a single-equation dynamic
regression model and is widely applicable to situations where the maturity
time of a forward contract exceeds the sampling period of the data. It (1)
avoids the need to use inefficient $HAC$\ robust inference and automatically
delivers consistent and asymptotically efficient estimates of the dynamic
regression parameters, (2) provides evidence on both short- and long-run
adjustments to the $UIP$ condition, and (3) facilitates incorporating
additional restrictions on the error process implied by the efficient
markets hypothesis ($EMH$) and rational expectations in foreign exchange
spot and forward markets.

$UIP$ asserts that the interest rate differential between two countries, or
equivalently the forward premium, is an efficient predictor of spot exchange
rate returns. This is requires the existence of rational expectations and a
constant risk premium. A widespread and important situation occurs when the
sampling frequency of the data exceeds the maturity time of the forward
contract. In this case the forward rate is a multi-step prediction of the
future spot rate, so the errors in $(UIP)$ regressions of spot returns on
forward premia are generally serially correlated.

Hansen and Hodrick (1980) noted that the consistency of the $GLS$ estimator
commonly invoked to correct for serial correlation requires strictly
exogenous regressors, which is unlikely to hold in $\mathit{UIP}$
regressions, and they therefore recommended using $OLS$\ with a $HAC$ robust
estimated covariance matrix. The $OLS$ estimator would then be a consistent,
albeit inefficient, estimator of the regression parameters, and this has led
to a plethora of $HAC$ regression methods (e.g., Newey and West, 1987).

An alternative approach has been to estimate a vector autoregression, ($VAR$%
) and then to test cross-equation restrictions that correspond to the $EMH$;
see Hakkio (1981), Baillie, Lippens and McMahon (1983) and Levy and Nobay
(1986). Some comparisons between the different methodologies are given in
Hodrick (1987)\ and Baillie (1989). The potential advantage with the $VAR$
approach is that it generally provides increased asymptotic efficiency
compared with the single equation approach. The disadvantage is that it
requires the specification and estimation of a full multi-equation $VAR.$

In this paper we propose a different approach, based on a single dynamic
regression, which we call $DynReg$. It requires only $OLS$ estimation and
does not require strict exogeneity, yet we show both theoretically and in
simulations that it is consistent and asymptotically efficient, and that
associated hypothesis tests have good finite-sample size and power.

We apply the $DynReg$ method to $32$ years of weekly data, regressing spot
returns on the lagged forward premium. We find clear rejections of the $UIP$%
\ hypothesis (a $UIP$ regression parameter, $\beta$, of unity), consistent
with the presence of time-varying risk premia, yet our coefficient estimates
are more reasonable than those of many earlier studies. In particular, they
are remarkably stable across currencies and all positive, whereas previous
studies often found large negative $\beta$'s. We also provide rolling $%
DynReg $ estimates, which indicate quite stable and relatively similar
estimated $\beta$ coefficients across time and currencies. Similar analysis
is also provided for the forward rate forecast error regressed on past
errors. These results are more volatile over time and include periods when
the $UIP$\ condition cannot be rejected.

The plan of the rest of the paper is as follows. Section 2 describes the
formulations of the $UIP$\ hypothesis, reviews previous econometric tests,
describes the $DynReg$ procedure, and shows to implement it in the context
of $UIP$ tests. Section 3 presents simulation evidence documenting the fine
performance of $DynReg$ estimates of $UIP$ regressions compared to $OLS/HAC$%
. Section 4 describes the results of a $DynReg$ $UIP$ analysis of several
floating exchange rates. Section 5 provides a brief conclusion.

\section{UIP and EMH}

Here we develop both economic and econometric aspects of uncovered interest
parity and the efficient markets hypothesis.

\subsection{Conceptual Formulations}

The natural logarithm of the spot exchange rate at time $t$ is denoted by $%
s_{t},$which is denominated in terms of the amount of foreign currency per
one numeraire US\ dollar. While $f_{t}$ is the natural logarithm of the
corresponding forward exchange rate at time $t$ with maturity time, or
forecast horizon, of $k\geq 1$. \ On denoting the domestic nominal interest
rate as $i_{t}$ and the corresponding foreign interest rate as $i_{t}^{\ast
} $, then the theory of Uncovered Interest Parity ($UIP$) implies that \ 
\begin{equation}
E_{t}(s_{t+k}-s_{t})=\left( i_{t}-i_{t}^{\ast }\right) ,
\end{equation}%
where $E_{t}$ represents the conditional expectation based on a sigma field
of information available at time $t$. Hence $UIP$\ requires the twin
assumptions of rational expectations and a constant or zero risk premium. \
Given the no arbitrage condition, Covered Interest Parity $(CIP)$ condition
implies that $\left( i_{t}-i_{t}^{\ast }\right) =\left( f_{t}-s_{t}\right) $
and will hold as an identity, and as an empirical matter CIP does indeed
hold almost exactly (Frenkel and Levich, 1975; Taylor, 1987). Hence the $UIP$%
\ condition in equation (1) is also frequently expressed as 
\begin{equation}
E_{t}(s_{t+k}-s_{t})=\left( f_{t}-s_{t}\right) .
\end{equation}%
The condition can be tested from the regression 
\begin{equation}
\left( s_{t+k}-s_{t}\right) =\alpha +\beta \left( f_{t}-s_{t}\right)
+u_{t+k},
\end{equation}%
so that the $k$ period rate of appreciation of the spot rate is predictable
from the forward premium. A test of $UIP$ or the $EMH$, is that $%
H_{0}:\alpha =0$ and $\beta =1$ and the error process is subject to the
restriction 
\begin{equation}
Cov\left( u_{t+k}u_{t+k-j}\right) =0\text{\ \ for \ }j>k.
\end{equation}
\ Bilson (1981) and Fama (1984) analyzed the $k=1$ case with the sampling
frequency matching the maturity time of the forward contract, so that a
natural test of $UIP$\ and $EMH$ was to estimate the regression \ 
\begin{equation}
\Delta s_{t+1}=\alpha +\beta \left( f_{t}-s_{t}\right) +u_{t+1},
\end{equation}%
where $UIP$\ implies that that $H_{0}:\alpha =0$ and $\beta =1$ and $u_{t+1}$
is a serially uncorrelated white noise process. It has been noted by Fama
(1984)\ and many subsequent studies that the estimated slope coefficient is
frequently $\beta <0.$ This implies a violation of $UIP$ with the country
with the higher rate of interest having an appreciating currency rather than
a depreciating currency; which is known as the Forward Premium Anomaly.

Another way of testing $UIP$\ is to express the condition as the forward
rate forecast error being unpredictable and to estimate the model \ 
\begin{equation}
\left( s_{t+k}-f_{t}\right) =\alpha +\beta \left( s_{t}-f_{t-k}\right)
+u_{t+k},
\end{equation}%
and to test $H_{0}:\alpha =0$ and $\beta =0$ \ and was tested by Hansen and
Hodrick (1980).

Early work by Frenkel (1977, 1979) tested the hypothesis $f_{t}=E_{t}s_{t+1}$
\ by estimating the regression 
\begin{equation}
s_{t+1}=\alpha +\beta f_{t}+u_{t+1},
\end{equation}%
and testing that $\alpha =0,$ $\beta =1$ and $u_{t+1}$ serially
uncorrelated. $\ $These early studies which used monthly data with $30$ day
forward rates so that the maturity time of the forward contract exactly
matched the sampling interval of the data, generally found that the $EMH$
could not be rejected. \ However, equation (3) is complicated by the fact
that the variables in question are non stationary. In particular, see
Baillie and Bollerslev (1989), Husted and Rush (1990) and Corbae and
Ouliaris (1988) who all found strong evidence that nominal spot and forward
rates are well represented as $I(1)$\ processes, which also appear to be
cointegrated with the forward premium $\left( s_{t}-f_{t}\right) $ being
stationary. \ Hence either equations (1) or (2) provide the natural economic
theory to be tested.

It was also realized that more powerful tests of $UIP$\ and the $EMH$ could
be obtained by using higher-frequency data where the maturity time of the
forward contract exceeds the sampling interval of the data; so that $k>1$.
This initially led to weekly data being used by Hansen and Hodrick (1980),
Hakkio (1981), Baillie, Lippens and McMahon (1983), bi weekly data in Hansen
and Hodrick (1983); and daily data in Baillie and Osterberg (1997). The
availability of higher frequency data then led to the development of a
variety of other testing procedures. Both the specifications of the tests
for $UIP$ and $EMH$\ in equations (3) and (6) provides the interesting
complication, given in equation (4) that a valid linear model for $u_{t+k}$
would be an $MA(k-1)$ process, with the possibility of additional forms of
non-linearity. The question now arises as how equations (3) and (6) should
be estimated.

\subsection{Econometric Tests}

Both equations (3) and (6) can be expressed as linear regressions, 
\begin{equation}
y_{t+k}=\alpha +\beta x_{t}+u_{t+k}.
\end{equation}%
The estimation of equation (3)\ proceeds by setting $\ y_{t+k}=\left(
s_{t+k}-s_{t}\right) $ and \ $x_{t}=\left( f_{t}-s_{t}\right) $, and the
estimation of equation (6) has $y{_{t+k}=}\left( {s}_{t+k}-f_{t}\right) $ \
and $x_{t}=y{_{t}=(s}_{t}-f_{t-k})$. In both cases the error process is
defined in equation (4), with the precise $MA$ representation to be given
later. Both models have overlapping data with $k>1$, and both have error
processes where weak exogeneity is not in doubt, because $E\left\{
x_{t}u_{t+k}\right\} =0$.\ \ 

However, as noted by Hansen and Hodrick (1980), consistency of time series
versions of $GLS$ techniques require the strict econometric exogeneity of
the $x$ process in equation (8), in the sense that $E(u_{t+k}\mid x_{t}$\ , $%
x_{t-1},x_{t+1},......)=0,$, so that $x$ is uncorrelated with all past and
future values of $u$. $GLS$\ estimation of $\beta $ implicitly filters the
data, which distorts orthogonality conditions and renders $GLS$ inconsistent
in the absence of strict exogeneity.

Because of the possible lack of strict exogeneity in equation (8), producing
inconsistency of $GLS$, Hansen and Hodrick (1980) recommend the use of $OLS$
rather than $GLS$. $OLS$ is consistent but inefficient when disturbances are
serially correlated, and the usual OLS standard errors are inconsistent. One
can, however, work out the correct standard error. In particular, the
consistent but asymptotically inefficient $OLS$ estimator is 
\begin{equation*}
\widehat{\beta _{OLS}}=\left( \sum_{t=1}^{T}x_{t}x_{t}^{/}\right)
^{-1}\left( \sum_{t=1}^{T}x_{t}y_{t+k}\right) ,
\end{equation*}%
with limiting distribution 
\begin{equation}
T^{1/2}(\ \widehat{\beta _{OLS}}-\beta )\rightarrow N(0,M),
\end{equation}%
where $\beta $ is the true value of $\beta $ and $M=Q^{-1}\Omega Q^{-1}$,
with \ 
\begin{equation}
Q=p\lim \left( T^{-1}\sum_{t=1}^{k}x_{t}x_{t}^{/}\right) =p\lim \left(
T^{-1}X^{/}X\right) .
\end{equation}%
The practical use of the above result depends on the estimated covariance
matrix of the error process, so that $\ $%
\begin{equation}
\widehat{M}=Q^{-1}\widehat{\Omega }Q^{-1}
\end{equation}
Hansen and Hodrick (1980) recommended estimating $\Omega $ by a $k$%
-dimensional band diagonal matrix, which would allow for an $MA(k-1)$ error
process. Subsequently there has been a vast literature focusing on the
estimation of \ $\Omega $,\ which then leads to the use of robust
(\textquotedblleft HAC") standard errors with OLS-estimated regression
parameters. The method of Newey and West (1987) has become particularly
influential.

It is worth noting that the above complications and considerations do not
arise in the $VAR$ approach where the hypothesis that one variable is a $k$%
-step-ahead prediction of another variable can be handed by a set of
non-linear restrictions on the $VAR$ parameters. \ However, we will not
pursue this issue here since this has been previously discussed by Baillie
(1989) and our aim in this paper is to focus on an alternative to the above
robustness approach in single equation estimation.

Before explaining an alternative single equation procedure that delivers
asymptotically efficient parameter estimates and tests (unlike OLS/HAC), we
first note some additional restrictions to the theory of $EMH.$ Due to the $%
k $-1 period overlap in sequential $k$-step-ahead forecasts, $u_{t+k}$ can
be expected to be an $MA(k-1)$ process, 
\begin{equation}
u_{t+k}=\varepsilon _{t+k}-\theta _{1}\ \varepsilon _{t+k-1}-....-\theta
_{k-1}\varepsilon _{t+1}=\theta (L)\varepsilon _{t+k},
\end{equation}%
where \ $\varepsilon _{t+k}$ \ are white noise and $\ \theta (L)=\left(
1-\theta _{1}L-....-\theta _{k-1}L^{k-1}\right) $. \ Following the standard
approach of previous literature in using weekly data, the forward rate is
generally measured on the Tuesday of each week and the spot rate on the
Thursday. This method of defining the data produces an average of $22$ days
in the forward contract, which implies a maturity time of four weeks and two
days, or $(22/5)$, or $4.40$ weeks. On assuming $k=4$ in equation (1) then $y%
{_{t+k}}$ \ in equation (7) would have an autocorrelation pattern of $\rho
_{1}=17/22=0.77$, and $\rho _{2}=12/22=0.55$, $\rho _{3}=7/22=0.32$, $\rho
_{4}=2/22=0.09$ and $\rho _{k}=0,$ for $k\geq 5.$ These population
autocorrelations imply a unique, invertible, $MA(4)$ process: 
\begin{equation}
\theta (L)=\left( 1+0.8366L+0.7728L^{2}+0.6863L^{3}+0.2577L^{4}\right) ,
\end{equation}%
with roots of $\left( 0.1909\pm 1.1724i\right) $ and $\left( -1.5266\pm
0.6575i\right)$, and autoregressive representation $\pi (L)=\theta (L)^{-1}$.

\subsection{The DynReg Approach}

An attractive alternative to the Hansen-Hodrick $OLS/HAC$ approach -- in
part because it delivers efficient as opposed to merely consistent parameter
estimates -- is a single-equation Dynamic Regression (\textquotedblleft $%
DynReg$") approach.\footnote{%
Here we give a basic sketch; for a more complete treatment see Baillie,
Diebold, Kapetanios and Kim (2022).} Consider the $UIP$ equation (3) from
the perspective of a vector process $\mathbf{z}_{t}=\left\{
y_{t},x_{t}\right\} $. $z$ is assumed to be covariance stationary with a
Wold Decomposition of 
\begin{equation}
\mathbf{z}_{t}=\sum_{k=0}^{\infty }\Psi _{k}\mathbf{w}_{t-k}\text{\ }
\end{equation}%
and a corresponding $VAR$ representation of 
\begin{equation}
\mathbf{z}_{t}=\sum_{k=1}^{\infty }\Pi _{k}\mathbf{z}_{t-k}+\mathbf{w}_{t},
\end{equation}%
where $\mathbf{\Psi }_{k}$ and $\mathbf{\Pi }_{k}$ are absolutely summable
sequences of non-stochastic $2x2$ matrices with $\mathbf{\Psi }_{0}=\mathbf{%
I.}$ \ It is further assumed that $E\left( \mathbf{w}_{t}\mid
Z_{t-1}^{w}\right) =0$ \ a.s. and $E\left( \mathbf{w}_{t}\mathbf{w}%
_{t}^{/}\mid Z_{t-1}^{w}\right) =\mathbf{\Omega }_{w}$ a.s. with $\left\vert 
\mathbf{\Omega }_{w}\right\vert >0$ \ and $\left\vert \left\vert \mathbf{%
\Omega }_{w}\right\vert \right\vert <\infty $ and$\ sup_{t}(\left\vert
\left\vert \mathbf{w}_{t}\right\vert \right\vert ^{4})<\infty $ with $%
Z_{t-1}^{w}$ being the $\sigma $ sigma field generated by $\left\{ \mathbf{w}%
_{s};s\leq t\right\} .$

A single equation of the $VAR$ in equation (15) can be conveniently
expressed as 
\begin{equation}
y_{t}=\sum_{j=1}^{p}\phi _{j}y_{t-j}+\sum_{i=1}^{k}\sum_{j=0}^{q}\beta
_{i,j}x_{i,t-j}+\varepsilon _{t},
\end{equation}%
or 
\begin{equation}
\phi (L)y_{t}=\sum_{i=1}^{k}\beta _{i}(L)x_{i,t}+\varepsilon _{t}.
\end{equation}%
This single equation is the dynamic regression ($DynReg$) of interest. Its
parameters are $\phi (L)=\left( 1-\phi _{1}L-...-\phi _{p}L^{p}\right) $ and 
$\ \beta _{i}(L)=\left( \beta _{i,0}-\beta _{i,1}L-...-\beta
_{i,q}L^{q}\right) $, and there are $\left( k+p+kp\right) $ parameters in
total\footnote{%
In the simulations and empirical applications of subsequent sections we set $%
p=q$ and select $p$ using the the Schwarz (1978) ($BIC$) criterion, $ln(%
\widehat{\sigma }_{u,h}^{2})+\left\{ k(1+p)+p\right\} T^{-1}\ln (T)$,} The
full set of parameters are denoted by $\theta ^{/}=\left( \phi _{1},..\phi
_{p},\beta _{1,0}..\beta _{1,q},\beta _{2,0},...\beta _{k,q}\ \right) $.

%; and the order $p$, of the dynamic regression is
%based on the standard assumption that $p=p_{T}\rightarrow \infty $ \ \ and $%
%\ p_{T}=o_{p}[T/\ln (T)]^{1/4}$ \ as $T\rightarrow \infty $ and is
%consistent with the dynamic regression being represented by a finite
%dimensional number of lags in the dynamic regression. Then, from Hannan and
%Deistler (1988), 

The $OLS$ estimates of the $DynReg$ parameters are denoted by $\widehat{%
\theta }$, and following the standard assumptions in Grenander (1981) and
Hannan and Deistler (1988), then as $T\rightarrow \infty $, we have 
\begin{equation*}
\widehat{\theta }=\theta +O_{p}(T^{-1/2}),
\end{equation*}%
and moreover that 
\begin{equation*}
T^{1/2}(\widehat{\theta }-\theta )\rightarrow N(0,Q^{-1}),
\end{equation*}%
where $\theta $ is the true value of the parameters and $Q$ is analogous to
the definition in equation (10) and is $\ $ 
\begin{equation}
Q=p\lim T^{-1}\sum_{t=1}^{k}z_{t}z_{t}^{/},
\end{equation}%
where $\ z_{t}^{/}=\left(
y_{t-1},..y_{t-p},x_{1,t},...x_{k,t},x_{1,t-1},...x_{k,t-p}\right) $.

Under the null hypothesis of $EMH$\ with Rational Expectations and constant
risk premium, we wish to estimate the model in equation (8), subject to the
restriction of having the $MA(k)$ error process defined in equation (12);
namely $u_{t}=\theta (L)\varepsilon _{t}$. \ The estimation of the general
model in equation (8) can be specialized to either the Fama regression in
equation (3), or the forward rate forecast error model in equation (6). On
premultipying through equation (8) by the filter $\theta (L)^{-1}$, we
obtain \ 
\begin{equation}
\left\{ \theta (L)^{-1}y_{t+k}\right\} =\alpha ^{\ast }+\beta \left\{ \theta
(L)^{-1}x_{t}\right\} +\varepsilon _{t+k}
\end{equation}%
where the new intercept is $\ \alpha ^{\ast }=$ $\alpha \theta (1)^{-1}$. \
The filtered explanatory variable is uncorrelated with current and future
innovations, $\varepsilon _{t+k}$, so that strict exogeneity is satisfied.
Then estimation of \ equation (19)\ by $OLS$\ will produce consistent and
asymptotically efficient estimates of the regression parameters. In practice
it is convenient to use the approximation $\ \theta (L)^{-1}\approx \pi (L)$
\ where $\pi (L)=\left( 1-\pi _{1}L-...-\pi _{p}L^{p}\right) $ and is a
polynomial in the lag operator of order $p$ and has all its roots lying
outside the unit circle.\footnote{%
Again, the choice of $p$ can based on $BIC$.} The $DynReg$ will then be 
\begin{equation}
\pi (L)y_{t+k}=\alpha ^{\ast }+\beta \pi (L)x_{t}+\varepsilon _{t+k}\ 
\end{equation}%
which is a restricted version of the general dynamic regression in equation
(16) and can also be estimated by restricted $OLS$ and now contains $(k+1)p$
parameters.

For the case of weekly data, with $k=4$, and from equation (13), then \ 
\begin{equation*}
\theta (L)=\left( 1+0.8366L+0.7728L^{2}+0.6863L^{3}+0.2577L^{4}\right)
\end{equation*}%
and on inverting $\theta (L)$ we find $\pi (L)$ such that $\pi _{1}=-0.84$, $%
\pi _{2}=-0.07$, $\pi _{3}=0.02$, $\pi _{4}=0.38,$ $\pi _{5}=0.08$, etc. and
the weights quickly decay to zero after eleven lags. The above restricted $%
DynReg$ model or $RDynReg$ model, can be contrasted with the unrestricted $%
DynReg$ in equation (16). Both the restricted and unrestricted dynamic
regressions are reported in the following simulation results and also the
tests of the $EMH$\ based on the estimated models. We also report Likelihood
Ratio ($LR$)\ tests to compare the $OLS$\ model with the $DynReg$\ and also
to compare the $DynReg$ model to the $RDynReg$ model, which is based on the
full set of $EMH$ restrictions.

\section{Simulation Results}

The simulation work was based on observed weekly spot exchange rates, and an
artificially generated error process from equation (13). Hence the
artificially generated forward rate is \ 
\begin{equation*}
f_{t}={s_{t+4}}-u_{t+4}
\end{equation*}%
\ \ and is generated to satisfy the null hypothesis of rational expectations
and a time invariant risk premium. The innovations $\varepsilon _{t}$ are
generated from an assumed $NID(0,\sigma ^{2})$ process, where from equation
(13) it can be seen that $\ 2.8345\sigma ^{2}=Var(u_{t})$, where the $%
Var(u_{t})$ is calculated for each currency from an initial forward premium
regression. The artificial forward rates are then used to construct $y_{t+k}$
and $x_{t}$ in equation (8). The weekly spot exchange rates were from
January $1989$ through April $2021$, for the six major currencies of
Australia, Canada, Japan, New Zealand. Switzerland and $UK$ against the
numeraire $US$\ dollar. The spot rates were recorded on the Thursday of each
week and realized $T=1,941$ observations and were obtained from Bloomberg.
Monte Carlo results for the unrestricted and restricted $DynReg$ are
presented in Table 1. \ The first six rows are from estimation by $OLS$ of
the traditional static regression in equation (6), with the first row
reporting conventional $OLS$\ robust standard errors; while the next five
rows use different $HAC$ covariance matrices. In order, the methods are:
Hansen and Hodrick (1980), Newey West (1987), Andrews (1991),
Kiefer-Vogelsang (2001)\ and finally the Equally Weighted Cosine ($EWC$)
method of Lazarus et al (2018).

The seventh and eighth rows of Table 1, in contrast, provide results from
using the $DynReg$ and $RDynReg$ approaches. The $DynReg$ method has an
unrestricted parameterization as in equation (16), while the $RDynReg$
method imposes the restrictions associated with $UIP$ \ In all the estimated
models the lag order, $p$, is selected by $BIC$ for each simulation
replication.

The $DynReg$ and $RDynReg$ estimators of $\beta$ clearly have substantially
reduced biases and $MSE$s. This result holds for all six simulation designs,
corresponding to the six different spot rates. Hence the inclusion of lagged
information in estimation makes a large difference compared with static HAC
estimation of equation (6).

Table 1 also presents estimates of the empirical test size, which is the
probability of rejecting the null hypothesis when it is true. $OLS$ clearly
has poor size properties, and all other test statistics offer massive
improvement, with $DynReg$ and $RDynReg$ faring slightly better than the HAC
alternatives.

Finally, the simulation results of Figures 1--3 we show size-corrected power
curves for sample sizes $T=250$, $T=500$ and $T=1,000$, respectively, for $%
\beta \in [-0.3, 0.3]$. $DynReg$ clearly dominates, for all six currencies.
The high $DynReg$ test power is a natural consequence of its higher
estimation efficiency.

In summary, Table 1 and Figures 1--3 clearly indicate that the $DynReg$ and $%
RDynReg$ methods improve on all competitors in all dimensions.

\section{Empirical Results for Six Currencies}

The above methodology was also implemented on the same weekly spot exchange
rate data between January $1989$ through April $2021$ and were complemented
with the \textit{actual} $30$ day forward rate data which was recorded on
the Tuesday of each week. This provides $T=1,941$ observations for each
bi-variate system for each of the six currencies. In practice, due to the
occurrence of holidays, religious festivals, and weekends, all of which
produce market closures, the length of time between a forward rate and its
corresponding spot rate in the data set is between $19$ and $25$ days.

There are several sets of results; each of which includes both $OLS/HAC$ and 
$DynReg$ estimation. Table 2 presents results for the model in equation (6),
where the forward rate forecast error is regressed on its lagged value. The $%
OLS/HAC$ results have positive but small values for the estimated $\beta $
with significant rejections of the $\beta =0$ null for all countries apart
from Switzerland. The $DynReg$ results uniformly do not reject the null.

The more interesting results appear in Table 3. They are based on the
classic Fama forward premium regression (3), which has more economic and
financial intuition. The $OLS/HAC$ $\beta $ estimates are between 0.07 for
Canada and -0.11 for Japan. Three of the currencies have an estimated $\beta
<0$, which is the case originally emphasized by Fama (1984). None of these
six estimated $\beta $ coefficients are significantly different from zero at
conventional levels. However, the results of $DynReg$ estimation indicate
long-run $\beta $ in the range of 0.24 to 0.31 while the restricted $RDynReg$
are in the range of 0.30 to 0.40 for all six currencies. Hence the $DynReg$
results all indicate significant risk premia but less than those of early
studies with monthly data where the $\beta <0.$

The appropriate Likelihood Ratio ($LR$)\ test statistic for the hypothesis
of $UIP$\ is denoted by $\lambda _{LR}$ and shows overwhelming rejections of
the $UIP$ and $EMH$ for all six currencies. Hence this indicates the
importance of information in the lagged forward rate errors, which is likely
due to time variation in the risk premium.

Some further insights into testing the $UIP$ condition are obtained by
estimating the above models with five years of observations in each rolling
sample. The results are reported graphically in Figures 4 through 7. Figures
4 and 5 show the estimates of $\beta $ from the forward rate forecast error
regressions. \ The $OLS$ estimates in the left hand panel are considerably
more jagged and rough than those of the long run beta estimated by $DynReg$
in the right hand set of panels. The estimates of long run $\beta $ do not
significantly depart from zero in any case. The 95\% confidence bands for $%
DynReg$ almost entirely contain the null in the Hansen-Hodrick model (i.e. $%
\beta =0$) for Australian Dollar, Canadian Dollar, Japanese Yen and New
Zealand Dollar (i.e. Figs 4 and 5). For Swiss Franc and UK Pound, the band
mostly covers the null value. This is clearly not the case with OLS.

The results for the Fama regression in (5), shown by Figs 6 and 7, are
particularly interesting and indicate considerable stability over the
rolling sample from the $DynReg$ estimates. These estimates are all positive
and are typically around 0.4 instead of the $\beta =1$ implied by $UIP$.
Given that the confidence bands for the $DynReg$ estimates in Figs 6 and 7
stay above zero for all six currencies, the estimates are statistically
different to zero for virtually all sets of rolling regressions, which is
not the case with $OLS$. Switzerland has slightly increased $\beta $ during
the financial crisis and is otherwise quite stable. New Zealand has a
slightly lower $\beta$ value than the other currencies. \ 

Hence there is considerable evidence that the $UIP$ condition needs to be
appended with risk premium terms, or possibly some measure of informational
inefficiency. Models with appropriate variables could potentially be
included in the modeling framework introduced in this paper.

\section{Conclusions}

This paper has suggested a new single-equation test for $UIP$ and $EMH$
based on $OLS$ estimation of a dynamic regression. The approach provides
consistent and asymptotically efficient parameter estimates, and is not
dependent on assumptions of strict exogeneity. This new approach has the
advantage of being asymptotically more efficient than the common approach of
using $HAC$ robust standard errors in the static forward premium regression.
The method also has advantages of showing dynamic effects of risk premia, or
other events that may lead to rejection of $UIP$\ and $EMH$. The empirical
results when spot returns are regressed on the lagged forward premium are
all positive and remarkably stable across currencies. \ \

\newpage

\vspace{-20cm} 
\begin{table}[tbp]
\caption{Performance of Tests of $\protect\beta=0$ in Dynamic Regressions
versus $OLS$ with robust standard errors in the Hansen-Hodrick (1980) model:}
\label{tab1}%
\begin{minipage}{\linewidth}
\begin{equation*}
s_{t+k}-f_{t}=\protect\alpha
+\beta(s_{t}-f_{t-k})+u_{t+k}. 
\end{equation*}
Under $EMH$ that $H_0:\,\alpha=0$ and $\beta=0$ and $Cov(u_{t+k},\,u_{t+k-j})=0$ for $j>k$,
\end{minipage}
\par
\begin{center}
\renewcommand{\arraystretch}{1.3} 
\begin{tabular}{cccccccccc}
\hline\hline
&  & \multicolumn{3}{c}{Australian Dollar} &  & \multicolumn{3}{c}{Canadian
Dollar} &  \\ \cline{3-5}\cline{7-9}
&  & Bias & MSE & Level &  & Bias & MSE & Level &  \\ \hline
OLS &  & -0.0054 & 0.0019 & 0.2630 &  & -0.0034 & 0.0018 & 0.2620 &  \\ 
OLS-HH &  & -- & -- & 0.0660 &  & -- & -- & 0.0560 &  \\ 
OLS-NW &  & -- & -- & 0.0810 &  & -- & -- & 0.0760 &  \\ 
OLS-Andrews &  & -- & -- & 0.0760 &  & -- & -- & 0.0720 &  \\ 
OLS-KV &  & -- & -- & 0.0520 &  & -- & -- & 0.0540 &  \\ 
OLS-EWC &  & -- & -- & 0.0590 &  & -- & -- & 0.0660 &  \\ 
$DynReg$ &  & -0.0004 & 0.0012 & 0.0590 &  & 0.0010 & 0.0011 & 0.0570 &  \\ 
$RDynReg$ &  & -0.0018 & 0.0006 & 0.0550 &  & -0.0004 & 0.0006 & 0.0460 & 
\\ \hline
&  & \multicolumn{3}{c}{Japanese Yen} &  & \multicolumn{3}{c}{New Zealand
Dollar} &  \\ \cline{3-5}\cline{7-9}
&  & Bias & MSE & Level &  & Bias & MSE & Level &  \\ \hline
OLS &  & -0.0054 & 0.0017 & 0.2470 &  & -0.0021 & 0.0021 & 0.2880 &  \\ 
OLS-HH &  & -- & -- & 0.0650 &  & -- & -- & 0.0580 &  \\ 
OLS-NW &  & -- & -- & 0.0790 &  & -- & -- & 0.0780 &  \\ 
OLS-Andrews &  & -- & -- & 0.0760 &  & -- & -- & 0.0740 &  \\ 
OLS-KV &  & -- & -- & 0.0540 &  & -- & -- & 0.0620 &  \\ 
OLS-EWC &  & -- & -- & 0.0680 &  & -- & -- & 0.0660 &  \\ 
$DynReg$ &  & -0.0017 & 0.0011 & 0.0450 &  & -0.0002 & 0.0012 & 0.0430 &  \\ 
$RDynReg$ &  & -0.0010 & 0.0006 & 0.0590 &  & 0.0004 & 0.0007 & 0.0550 &  \\ 
\hline
\end{tabular}%
\end{center}
\end{table}

\newpage
\setcounter{table}{0}

\vspace{-80cm} 
\begin{table}[tbp]
\caption{(Continued) Performance of Tests of $\protect\beta=0$ in Dynamic
Regressions versus $OLS$ with robust standard errors in the Hansen-Hodrick
(1980) model:}
\label{tab1}\ContinuedFloat
\begin{minipage}{\linewidth}
%Performance of Tests of $\beta=0$ in Dynamic Regressions versus $OLS$ with robust standard errors in the model:
\begin{equation*}
s_{t+k}-f_{t}=\protect\alpha
+\beta(s_{t}-f_{t-k})+u_{t+k}. 
\end{equation*}
Under $EMH$ that $H_0:\,\alpha=0$ and $\beta=0$ and $Cov(u_{t+k},\,u_{t+k-j})=0$ for $j>k$,
\end{minipage}
%}
\par
\begin{center}
\renewcommand{\arraystretch}{1.3} 
\begin{tabular}{cccccccccc}
\hline\hline
&  & \multicolumn{3}{c}{Swiss Franc} &  & \multicolumn{3}{c}{UK Pound} &  \\ 
\cline{3-5}\cline{7-9}
&  & Bias & MSE & Level &  & Bias & MSE & Level &  \\ \hline
OLS &  & -0.0047 & 0.0017 & 0.2420 &  & -0.0029 & 0.0018 & 0.2350 &  \\ 
OLS-HH &  & -- & -- & 0.0640 &  & -- & -- & 0.0650 &  \\ 
OLS-NW &  & -- & -- & 0.0720 &  & -- & -- & 0.0740 &  \\ 
OLS-Andrews &  & -- & -- & 0.0710 &  & -- & -- & 0.0740 &  \\ 
OLS-KV &  & -- & -- & 0.0440 &  & -- & -- & 0.0530 &  \\ 
OLS-EWC &  & -- & -- & 0.0610 &  & -- & -- & 0.0660 &  \\ 
$DynReg$ &  & -0.0001 & 0.0011 & 0.0470 &  & -0.0008 & 0.0010 & 0.0390 &  \\ 
$RDynReg$ &  & -0.0015 & 0.0006 & 0.0480 &  & -0.0003 & 0.0006 & 0.0460 & 
\\ \hline
\multicolumn{10}{c}{\begin{minipage}[t]{0.9\columnwidth}{Key: The first six
test statistics are based on $OLS$ estimation of $\beta$ with standard
errors based on estimated parameters estimated covariance matrix computed by
(i) regular $OLS$, (ii) $HH$, method of Hansen and Hodrick, (iii) $NW$,
method of Newey and West, (iv) $Andrews$, method of Andrews, (v) $KV$,
method Kiefer and Vogelsang, (vi) $EWC$, Equally Weighted Cosine. The
$DynReg$ statistics are based on estimation of the unrestricted dynamic
regression in equation (16) and $RDynReg$ is the restricted dynamic
regression in equation (19) that constrains the error to be an $MA$(k-1)
process as defined in equation (12). }\end{minipage}}%
\end{tabular}%
\end{center}
\end{table}

\newpage

\begin{figure}[tbp]
\centering
\begin{tabular}{ccc}
\setlength{\itemsep}{-1.0cm} \psfig{file = 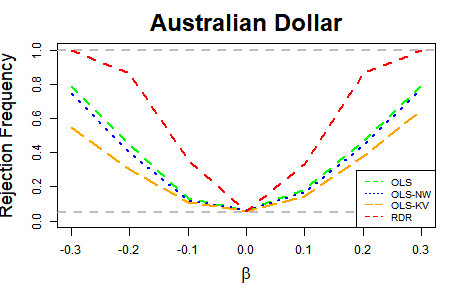, width = 7cm, angle=0}
& \psfig{file = 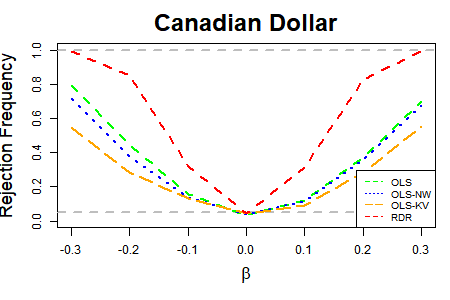, width = 7cm, angle=0} &  \\ 
\psfig{file = 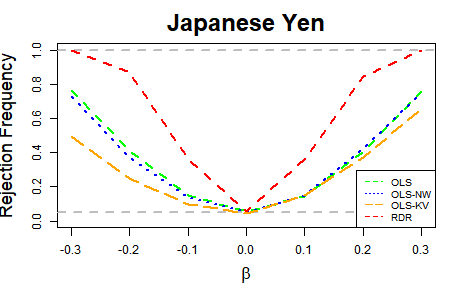, width = 7cm, angle=0} & \psfig{file = 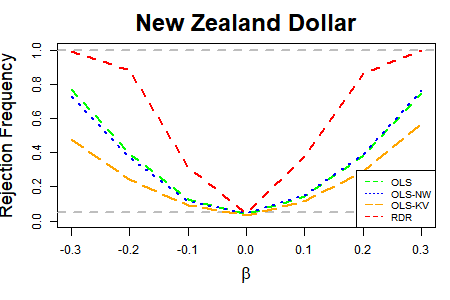,
width = 7cm, angle=0} &  \\ 
\psfig{file = 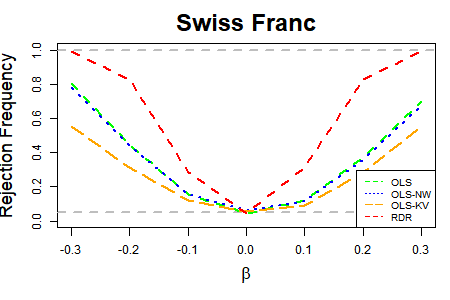, width = 7cm, angle=0} & \psfig{file = 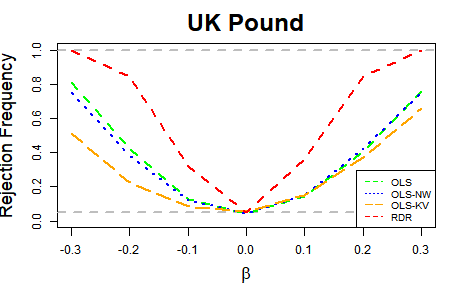,
width = 7cm, angle=0} &  \\ 
&  & 
\end{tabular}
{\small {} }
\caption{Size-corrected power for OLS (green), OLS-NW (blue), OLS-KV
(yellow) and $RDynReg$ (red): A 5\% t-test is conducted for the sample size 
\textbf{T=250}. The underlying model is the \textit{Hansen-Hodrick} model
(1980). The null value is $\protect\beta=0$ and the alternatives are $%
\protect\beta=\pm0.1, \pm0.2, \pm0.3$.}
\label{fig:rates}
\end{figure}

\newpage

\begin{figure}[tbp]
\centering
\begin{tabular}{ccc}
\setlength{\itemsep}{-1.0cm} \psfig{file = 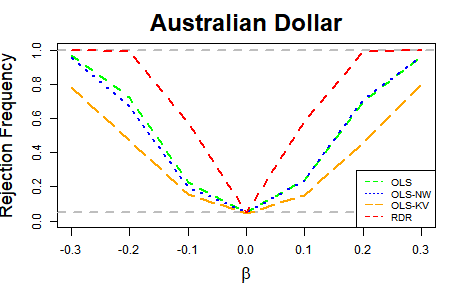, width = 7cm, angle=0}
& \psfig{file = 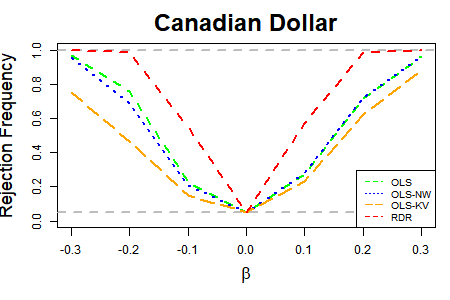, width = 7cm, angle=0} &  \\ 
\psfig{file = 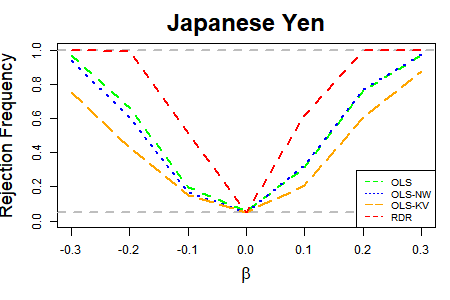, width = 7cm, angle=0} & \psfig{file = 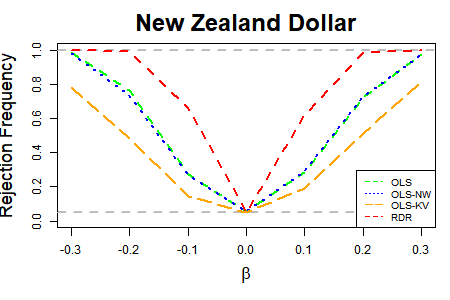,
width = 7cm, angle=0} &  \\ 
\psfig{file = 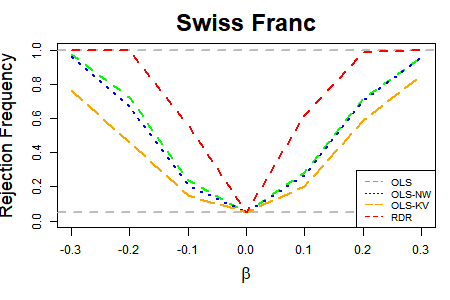, width = 7cm, angle=0} & \psfig{file = 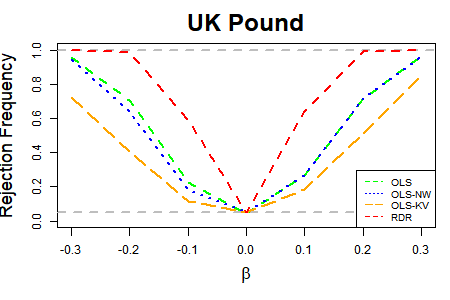,
width = 7cm, angle=0} &  \\ 
&  & 
\end{tabular}
{\small {} }
\caption{Size-corrected power for OLS (green), OLS-NW (blue), OLS-KV
(yellow) and $RDynReg$ (red): A 5\% t-test is conducted for the sample size 
\textbf{T=500}. The underlying model is the \textit{Hansen-Hodrick} model
(1980). The null value is $\protect\beta=0$ and the alternatives are $%
\protect\beta=\pm0.1, \pm0.2, \pm0.3$.}
\label{fig:rates}
\end{figure}

\newpage

\begin{figure}[tbp]
\centering
\begin{tabular}{ccc}
\setlength{\itemsep}{-1.0cm} \psfig{file = 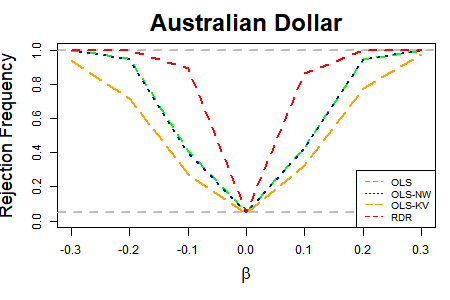, width = 7cm, angle=0}
& \psfig{file = 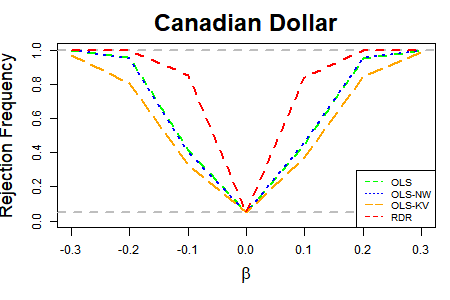, width = 7cm, angle=0} &  \\ 
\psfig{file = 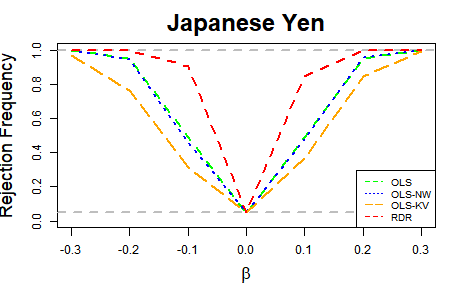, width = 7cm, angle=0} & \psfig{file =
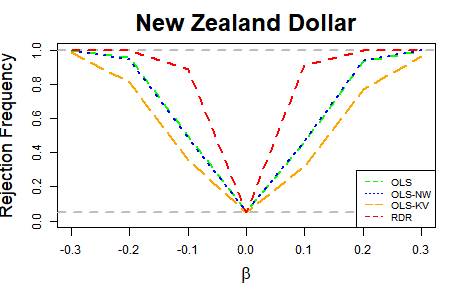, width = 7cm, angle=0} &  \\ 
\psfig{file = 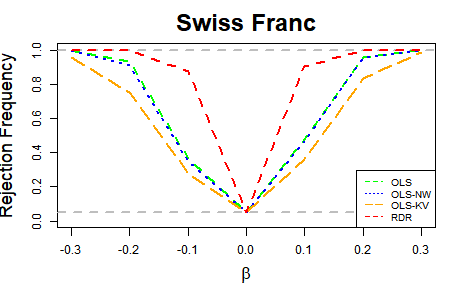, width = 7cm, angle=0} & \psfig{file =
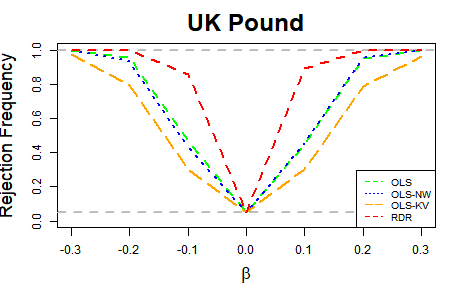, width = 7cm, angle=0} &  \\ 
&  & 
\end{tabular}
{\small {} }
\caption{Size-corrected power for OLS (green), OLS-NW (blue), OLS-KV
(yellow) and $RDynReg$ (red): A 5\% t-test is conducted for the sample size 
\textbf{T=1000}. The underlying model is the \textit{Hansen-Hodrick} (1980)
model. The null value is $\protect\beta=0$ and the alternatives are $\protect%
\beta=\pm0.1, \pm0.2, \pm0.3$.}
\label{fig:rates}
\end{figure}

\vspace{-20cm} 
\begin{table}[tbp]
\caption{Estimation of the Hansen-Hodrick (1980) model $s_{t+k}-f_{t}=%
\protect\alpha+\protect\beta(s_{t}-f_{t-k})+u_{t+k}$ for actual weekly spot
and forward exchange rate data from January 1989 through April 2021. }
\label{tab1}
\begin{center}
\renewcommand{\arraystretch}{1.3} 
\begin{tabular}{cccccccccc}
\hline\hline
&  & \multicolumn{3}{c}{Australian Dollar} &  & \multicolumn{3}{c}{Canadian
Dollar} &  \\ \cline{3-5}\cline{7-9}
&  & $\hat{\beta}$ & $s.e.(\hat{\beta})$ & $\lambda_{LR}$ &  & $\hat{\beta}$
& $s.e.(\hat{\beta})$ & $\lambda_{LR}$ &  \\ \hline
OLS &  & 0.1393 & 0.0241 & -- &  & 0.1210 & 0.0243 & -- &  \\ 
OLS-HH &  & -- & 0.0664 & -- &  & -- & 0.0536 & -- &  \\ 
OLS-NW &  & -- & 0.0606 & -- &  & -- & 0.0488 & -- &  \\ 
OLS-Andrews &  & -- & 0.0620 & -- &  & -- & 0.0479 & -- &  \\ 
OLS-KV &  & -- & 0.0158 & -- &  & -- & 0.0069 & -- &  \\ 
OLS-EWC &  & -- & 0.0566 & -- &  & -- & 0.0448 & -- &  \\ 
$DynReg$ &  & -0.0776 & 0.0320 & 2408.6 &  & -0.0167 & 0.0320 & 2226.6 &  \\ 
$RDynReg$ &  & 0.0024 & 0.0245 & -- &  & -0.0003 & 0.0246 & -- &  \\ \hline
&  & \multicolumn{3}{c}{Japanese Yen} &  & \multicolumn{3}{c}{New Zealand
Dollar} &  \\ \cline{3-5}\cline{7-9}
&  & $\hat{\beta}$ & $s.e.(\hat{\beta})$ & $\lambda_{LR}$ &  & $\hat{\beta}$
& $s.e.(\hat{\beta})$ & $\lambda_{LR}$ &  \\ \hline
OLS &  & 0.1503 & 0.0242 & -- &  & 0.0955 & 0.0255 & -- &  \\ 
OLS-HH &  & -- & 0.0533 & -- &  & -- & 0.0598 & -- &  \\ 
OLS-NW &  & -- & 0.0483 & -- &  & -- & 0.0566 & -- &  \\ 
OLS-Andrews &  & -- & 0.0486 & -- &  & -- & 0.0582 & -- &  \\ 
OLS-KV &  & -- & 0.0098 & -- &  & -- & 0.0089 & -- &  \\ 
OLS-EWC &  & -- & 0.0497 & -- &  & -- & 0.0571 & -- &  \\ 
$DynReg$ &  & 0.0120 & 0.0324 & 2479.5 &  & -0.0555 & 0.0338 & 2386.3 &  \\ 
$RDynReg$ &  & 0.0192 & 0.0245 & -- &  & -0.0236 & 0.0258 & -- &  \\ \hline
&  & \multicolumn{3}{c}{Swiss Franc} &  & \multicolumn{3}{c}{UK Pound} &  \\ 
\cline{3-5}\cline{7-9}
&  & $\hat{\beta}$ & $s.e.(\hat{\beta})$ & $\lambda_{LR}$ &  & $\hat{\beta}$
& $s.e.(\hat{\beta})$ & $\lambda_{LR}$ &  \\ \hline
OLS &  & 0.0373 & 0.0244 & -- &  & 0.0799 & 0.0244 & -- &  \\ 
OLS-HH &  & -- & 0.0474 & -- &  & -- & 0.0598 & -- &  \\ 
OLS-NW &  & -- & 0.0446 & -- &  & -- & 0.0540 & -- &  \\ 
OLS-Andrews &  & -- & 0.0455 & -- &  & -- & 0.0542 & -- &  \\ 
OLS-KV &  & -- & 0.0333 & -- &  & -- & 0.0158 & -- &  \\ 
OLS-EWC &  & -- & 0.0438 & -- &  & -- & 0.0576 & -- &  \\ 
$DynReg$ &  & -0.0415 & 0.0322 & 2315.3 &  & -0.0527 & 0.0321 & 2391.8 &  \\ 
$RDynReg$ &  & 0.0280 & 0.0245 & -- &  & -0.0584 & 0.0245 & -- &  \\ \hline
\multicolumn{10}{c}{\begin{minipage}[t]{0.9\columnwidth}{Key: See key to
Table 1. The Likelihood Ratio test statistic $\lambda_{LR}$ is from a test
of the static $OLS$ regression model against the DynReg model.}
\end{minipage}}%
\end{tabular}%
\end{center}
\end{table}

\vspace{-20cm} 
\begin{table}[tbp]
\caption{Estimation of the Fama (1984) model $s_{t+k}-s_{t}=\protect\alpha+%
\protect\beta(f_{t}-s_{t})+u_{t+k}$ for actual weekly spot and forward
exchange rate data from January 1989 through April 2021. }
\label{tab1}
\begin{center}
\renewcommand{\arraystretch}{1.3} 
\begin{tabular}{cccccccccc}
\hline\hline
&  & \multicolumn{3}{c}{Australian Dollar} &  & \multicolumn{3}{c}{Canadian
Dollar} &  \\ \cline{3-5}\cline{7-9}
&  & $\hat{\beta}$ & $s.e.(\hat{\beta})$ & $\lambda_{LR}$ &  & $\hat{\beta}$
& $s.e.(\hat{\beta})$ & $\lambda_{LR}$ &  \\ \hline
OLS &  & 0.0115 & 0.0815 & -- &  & 0.0727 & 0.0762 & -- &  \\ 
OLS-HH &  & -- & 0.1077 & -- &  & -- & 0.1124 & -- &  \\ 
OLS-NW &  & -- & 0.1055 & -- &  & -- & 0.1141 & -- &  \\ 
OLS-Andrews &  & -- & 0.1043 & -- &  & -- & 0.1129 & -- &  \\ 
OLS-KV &  & -- & 0.0300 & -- &  & -- & 0.0270 & -- &  \\ 
OLS-EWC &  & -- & 0.1068 & -- &  & -- & 0.1057 & -- &  \\ 
$DynReg$ &  & 0.2440 & 0.0503 & 2852.4 &  & 0.3038 & 0.0495 & 2669.7 &  \\ 
$RDynReg$ &  & 0.3696 & 0.0340 & -- &  & 0.3972 & 0.0325 & -- &  \\ \hline
&  & \multicolumn{3}{c}{Japanese Yen} &  & \multicolumn{3}{c}{New Zealand
Dollar} &  \\ \cline{3-5}\cline{7-9}
&  & $\hat{\beta}$ & $s.e.(\hat{\beta})$ & $\lambda_{LR}$ &  & $\hat{\beta}$
& $s.e.(\hat{\beta})$ & $\lambda_{LR}$ &  \\ \hline
OLS &  & -0.1148 & 0.0815 & -- &  & -0.0320 & 0.0842 & -- &  \\ 
OLS-HH &  & -- & 0.0971 & -- &  & -- & 0.0847 & -- &  \\ 
OLS-NW &  & -- & 0.0927 & -- &  & -- & 0.0898 & -- &  \\ 
OLS-Andrews &  & -- & 0.0907 & -- &  & -- & 0.0902 & -- &  \\ 
OLS-KV &  & -- & 0.0171 & -- &  & -- & 0.0639 & -- &  \\ 
OLS-EWC &  & -- & 0.0842 & -- &  & -- & 0.0877 & -- &  \\ 
$DynReg$ &  & 0.2750 & 0.0503 & 2847.1 &  & 0.2392 & 0.0514 & 2845.9 &  \\ 
$RDynReg$ &  & 0.3462 & 0.0334 & -- &  & 0.3398 & 0.0351 & -- &  \\ \hline
&  & \multicolumn{3}{c}{Swiss Franc} &  & \multicolumn{3}{c}{UK Pound} &  \\ 
\cline{3-5}\cline{7-9}
&  & $\hat{\beta}$ & $s.e.(\hat{\beta})$ & $\lambda_{LR}$ &  & $\hat{\beta}$
& $s.e.(\hat{\beta})$ & $\lambda_{LR}$ &  \\ \hline
OLS &  & 0.0277 & 0.0770 & -- &  & -0.0737 & 0.0861 & -- &  \\ 
OLS-HH &  & -- & 0.1545 & -- &  & -- & 0.1066 & -- &  \\ 
OLS-NW &  & -- & 0.1491 & -- &  & -- & 0.1075 & -- &  \\ 
OLS-Andrews &  & -- & 0.1489 & -- &  & -- & 0.1058 & -- &  \\ 
OLS-KV &  & -- & 0.0983 & -- &  & -- & 0.0725 & -- &  \\ 
OLS-EWC &  & -- & 0.1486 & -- &  & -- & 0.1142 & -- &  \\ 
$DynReg$ &  & 0.2427 & 0.0477 & 2796.6 &  & 0.3054 & 0.0503 & 2797.2 &  \\ 
$RDynReg$ &  & 0.3012 & 0.0326 & -- &  & 0.3566 & 0.0365 & -- &  \\ \hline
\multicolumn{10}{c}{\begin{minipage}[t]{0.9\columnwidth}{Key: See key to
Table 1. The Likelihood Ratio test statistic $\lambda_{LR}$ is from a test
of the static $OLS$ regression model against the $DynReg$
model.}\end{minipage}}%
\end{tabular}%
\end{center}
\end{table}

\begin{figure}[tbp]
\centering
\begin{tabular}{ccc}
\setlength{\itemsep}{-1.0cm} \psfig{file = 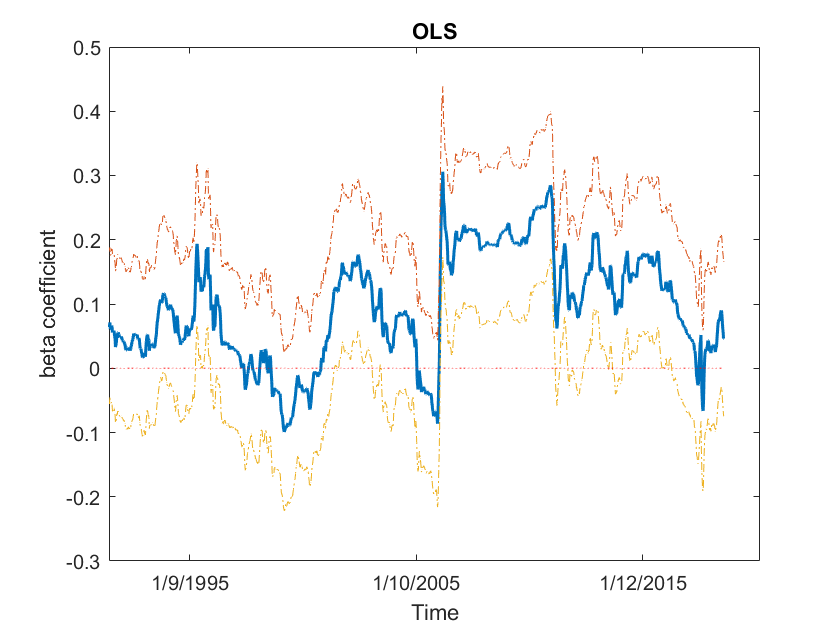, width = 7cm,
angle=0} & \psfig{file = 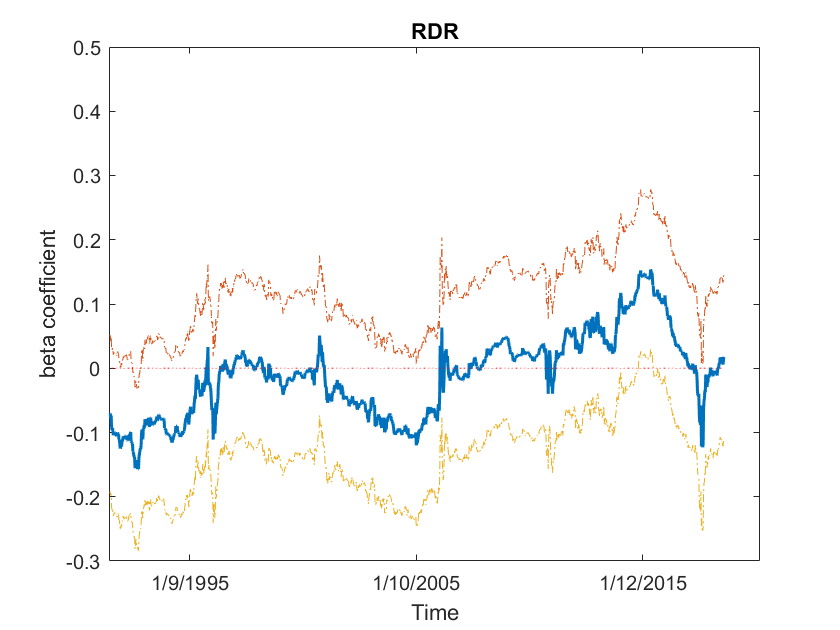, width = 7cm, angle=0} &  \\ 
Australian Dollar & Australian Dollar &  \\ 
\psfig{file = 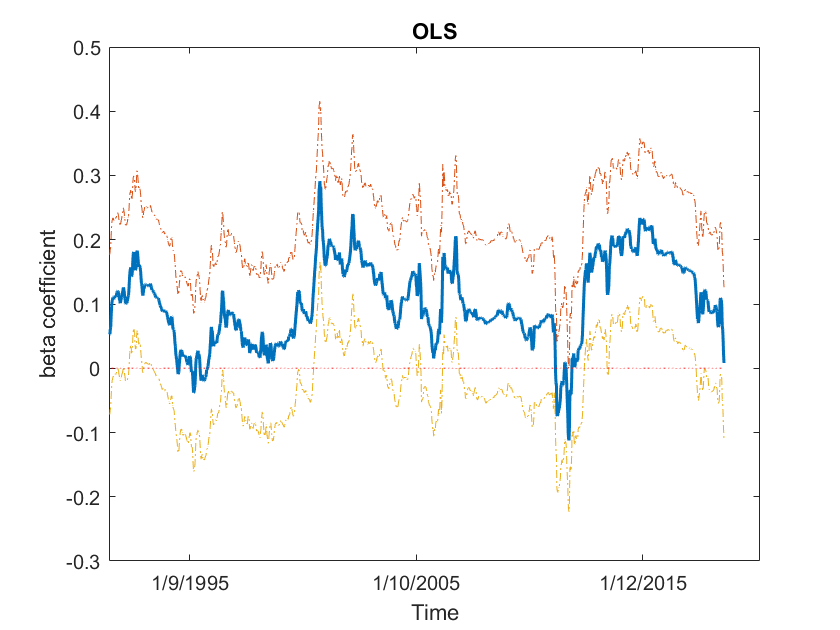, width = 7cm, angle=0} & \psfig{file =
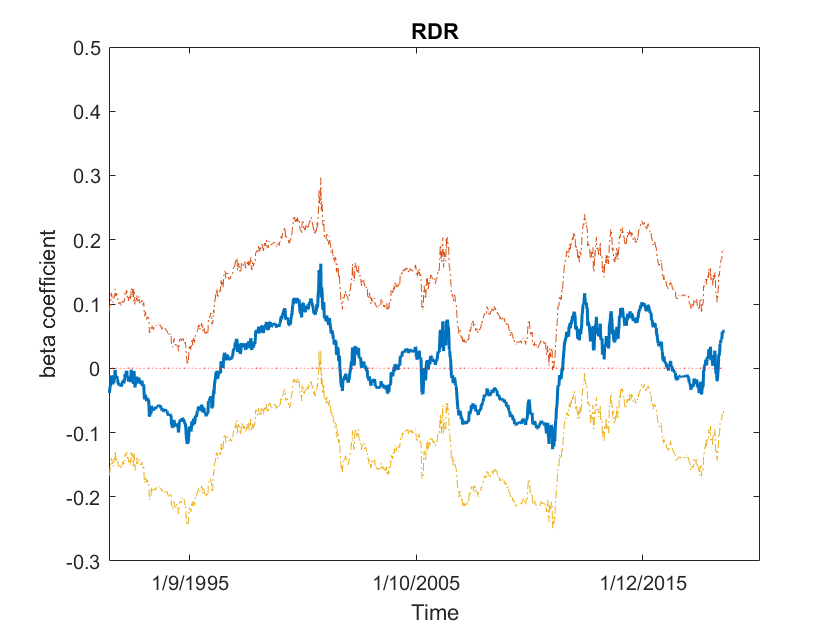, width = 7cm, angle=0} &  \\ 
Canadian Dollar & Canadian Dollar &  \\ 
\psfig{file = 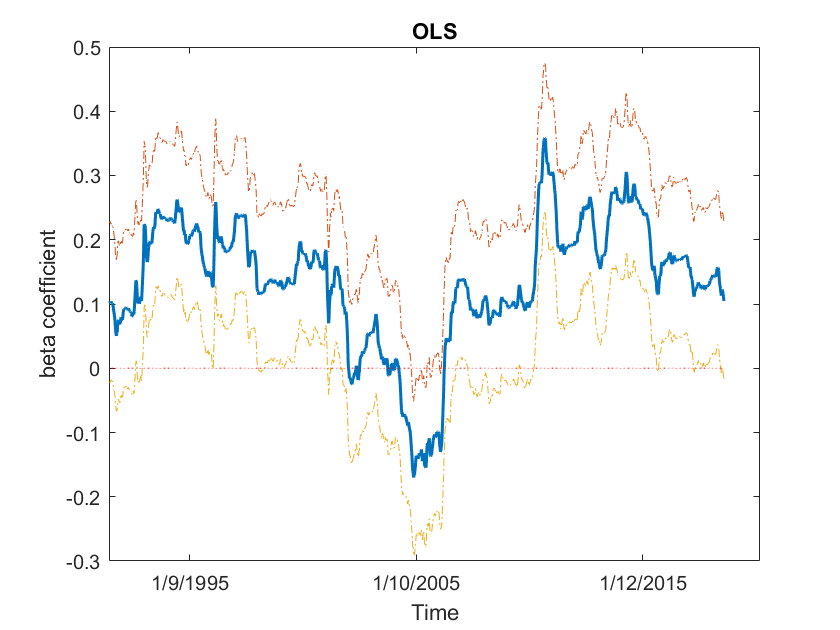, width = 7cm, angle=0} & \psfig{file =
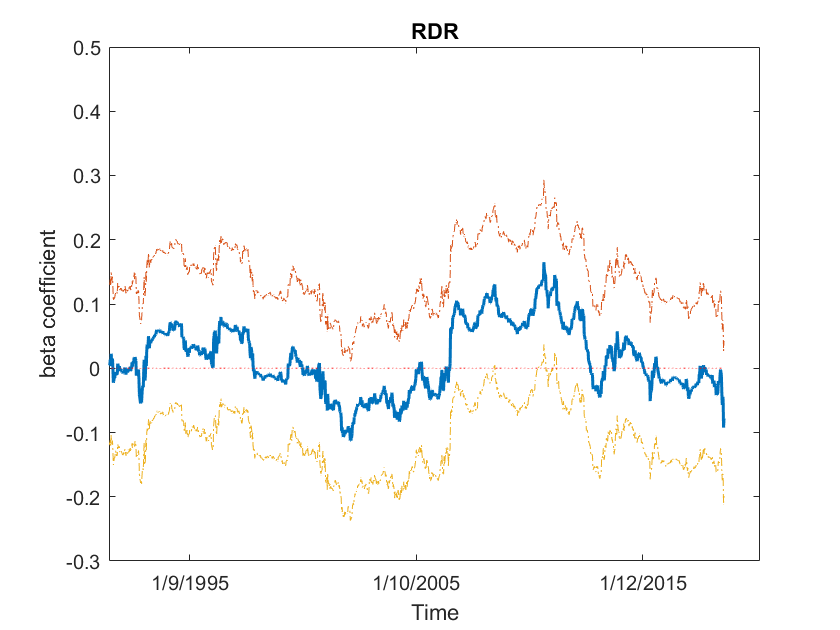, width = 7cm, angle=0} &  \\ 
Japanese Yen & Japanese Yen &  \\ 
&  & 
\end{tabular}
{\small {} }
\caption{The blue curve is the 5-year rolling OLS (left)/$RDynReg$ (right)
estimate of $\protect\beta$ in the \textit{Hansen-Hodrick} (1980) model; The
null is $\protect\beta=0$ and the dashed ones are the 95\% confidence bands.}
\label{fig:rates}
\end{figure}

\begin{figure}[tbp]
\centering
\begin{tabular}{ccc}
\setlength{\itemsep}{-1.0cm} \psfig{file = 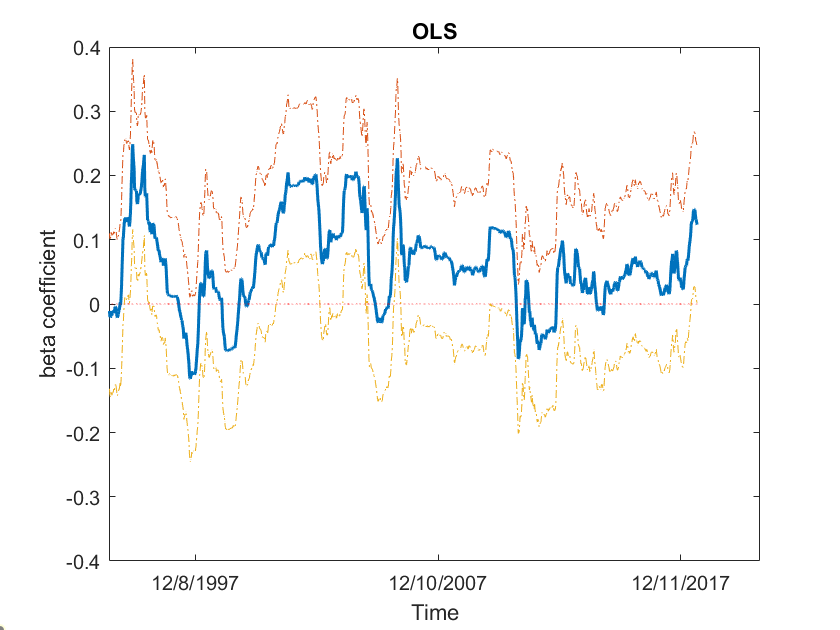, width = 7cm,
angle=0} & \psfig{file = 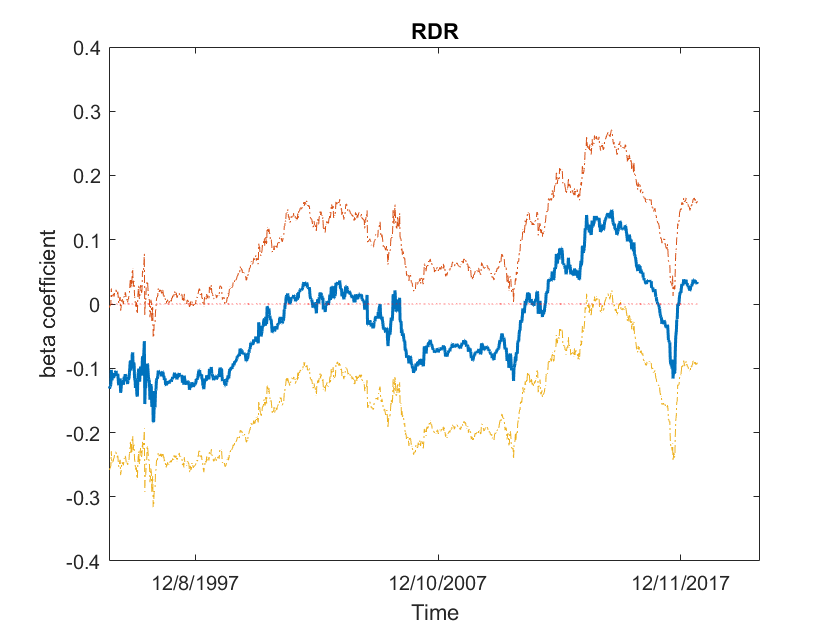, width = 7cm, angle=0} &  \\ 
New Zealand Dollar & New Zealand Dollar &  \\ 
\psfig{file = 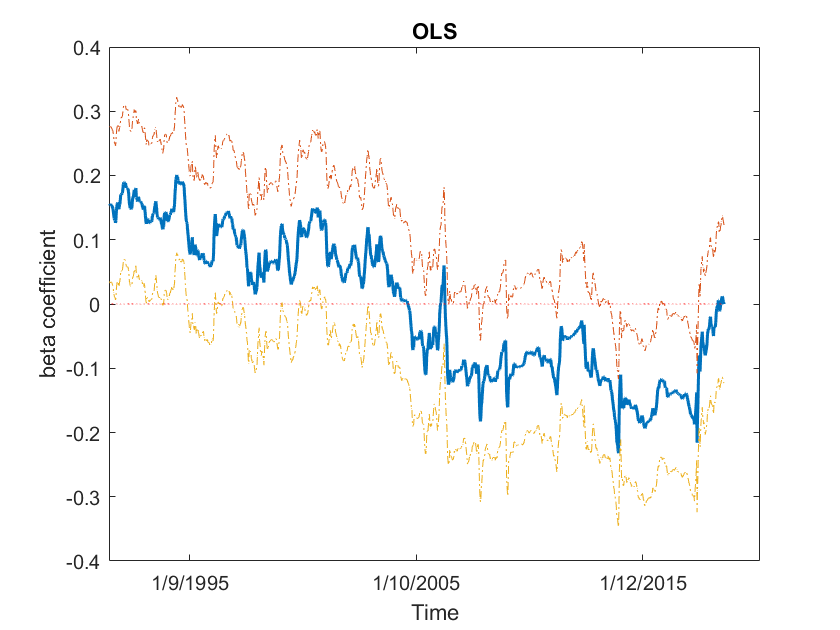, width = 7cm, angle=0} & \psfig{file =
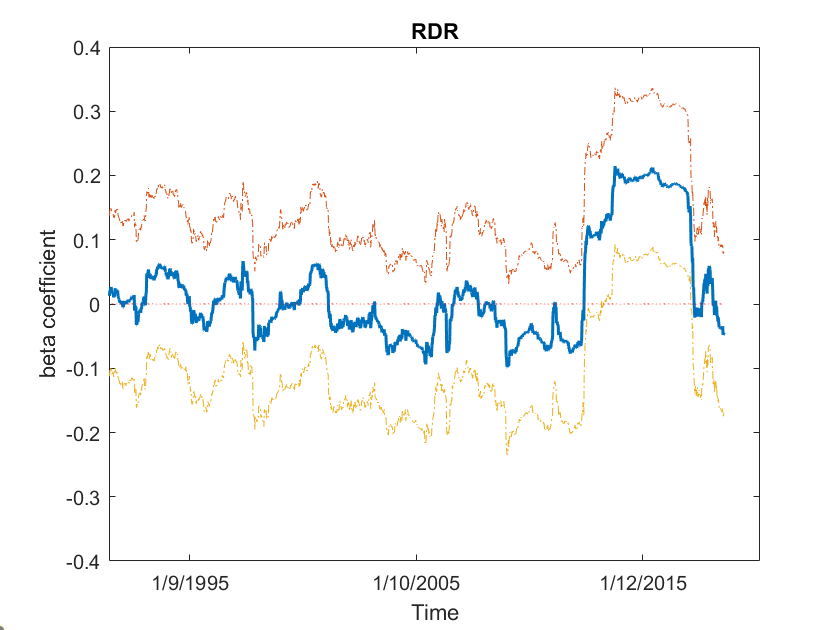, width = 7cm, angle=0} &  \\ 
Swiss Franc & Swiss Franc &  \\ 
\psfig{file = 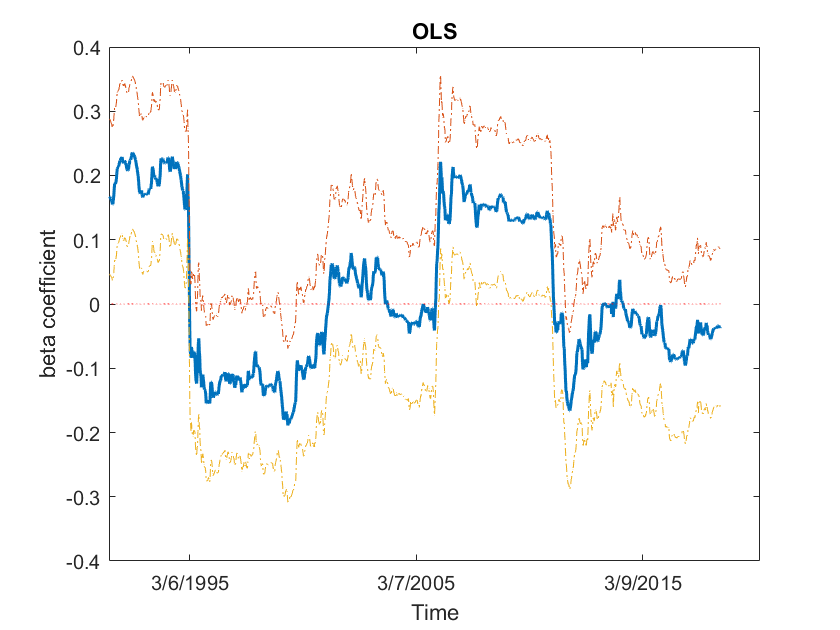, width = 7cm, angle=0} & \psfig{file =
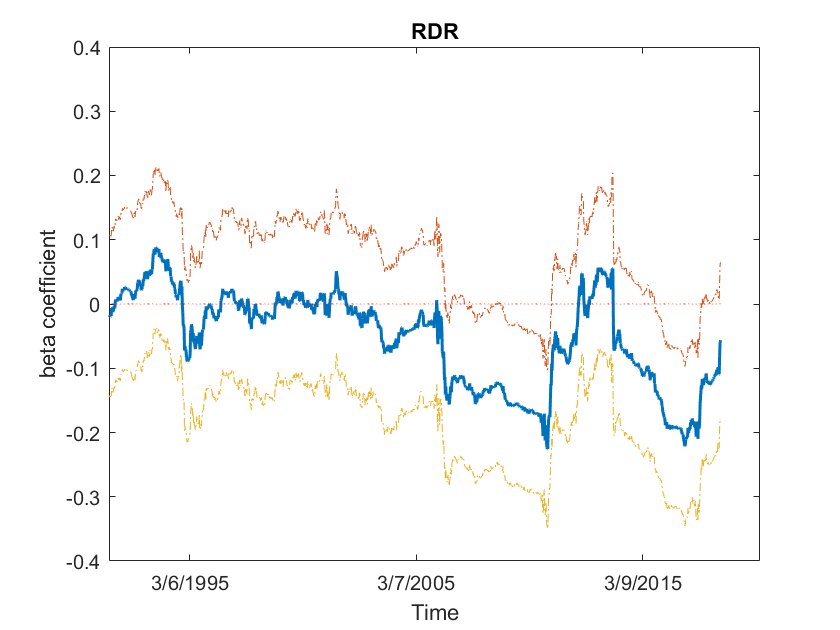, width = 7cm, angle=0} &  \\ 
UK Pound & UK Pound &  \\ 
&  & 
\end{tabular}
{\small {} }
\caption{The blue curve is the 5-year rolling OLS (left)/$RDynReg$ (right)
estimate of $\protect\beta$ in the \textit{Hansen-Hodrick} (1980) model; The
null is $\protect\beta=0$ and the dashed ones are the 95\% confidence bands.}
\label{fig:rates}
\end{figure}

\begin{figure}[tbp]
\centering
\begin{tabular}{ccc}
\setlength{\itemsep}{-1.0cm} \psfig{file = 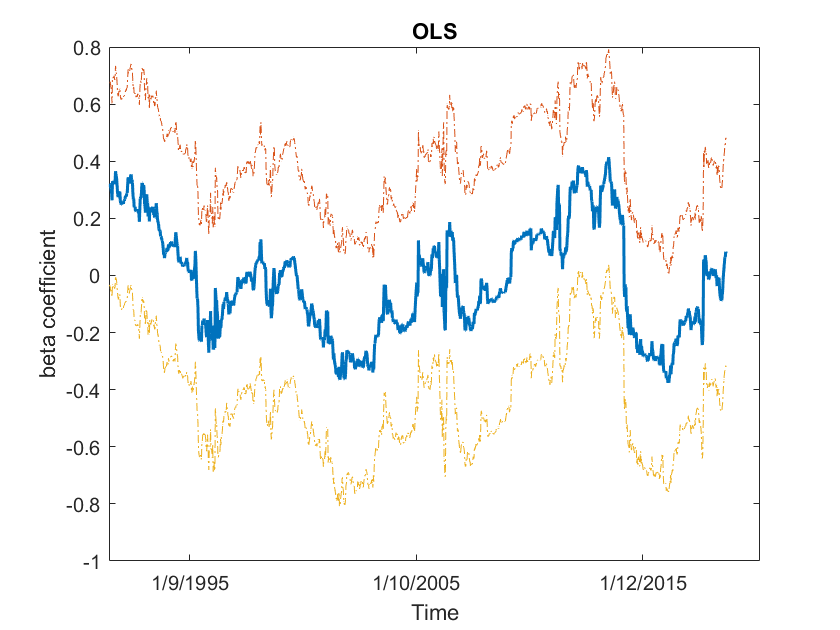, width = 7cm,
angle=0} & \psfig{file = 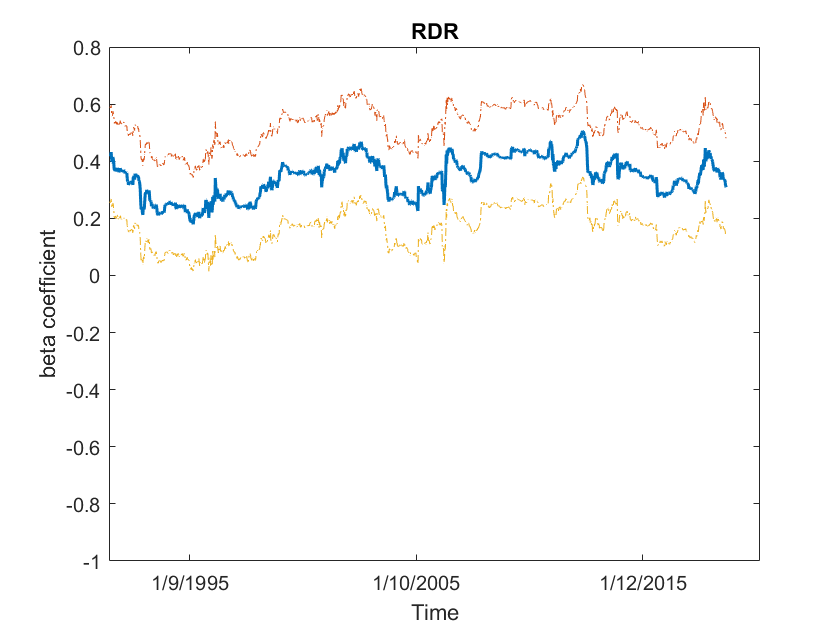, width = 7cm, angle=0} &  \\ 
Australian Dollar & Australian Dollar &  \\ 
\psfig{file = 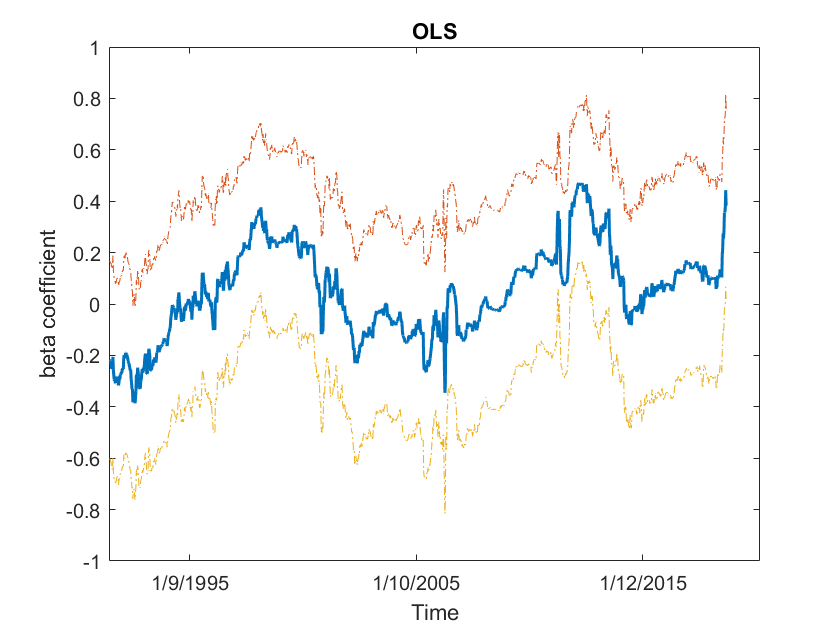, width = 7cm, angle=0} & \psfig{file =
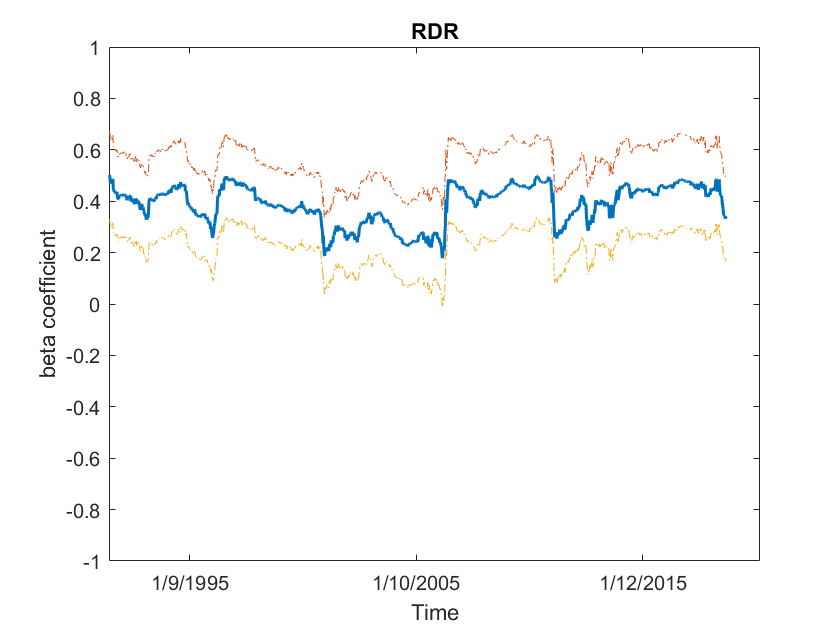, width = 7cm, angle=0} &  \\ 
Canadian Dollar & Canadian Dollar &  \\ 
\psfig{file = 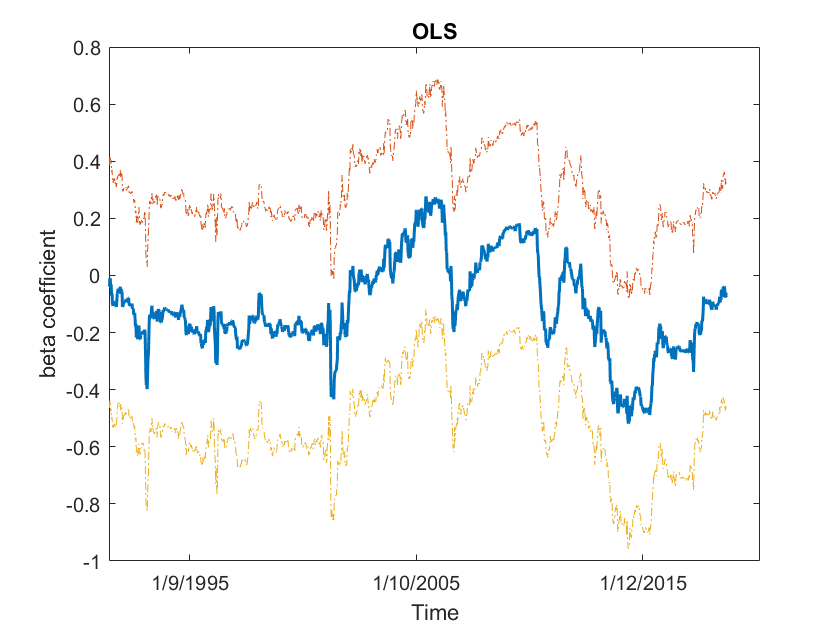, width = 7cm, angle=0} & \psfig{file =
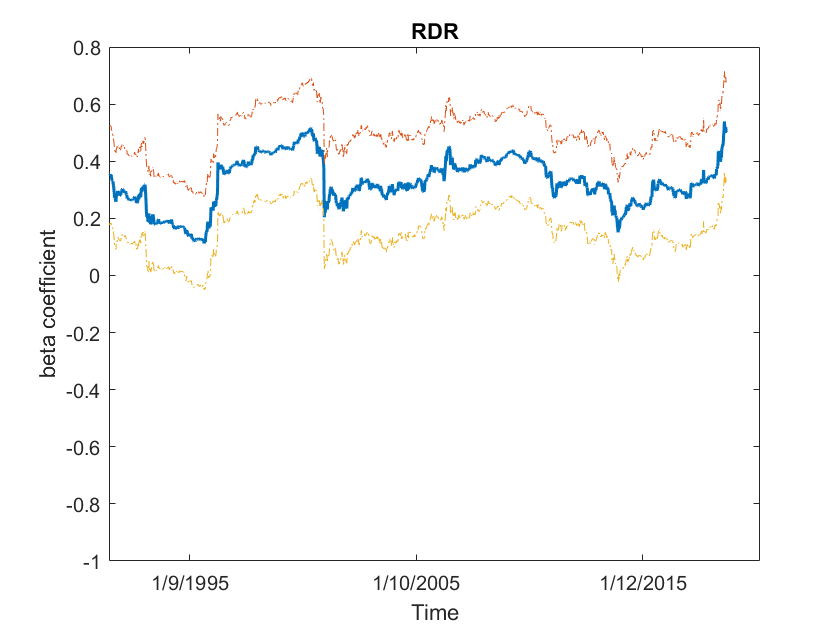, width = 7cm, angle=0} &  \\ 
Japanese Yen & (f) Japanese Yen &  \\ 
&  & 
\end{tabular}
{\small {} }
\caption{The blue curve is the 5-year rolling OLS (left)/$RDynReg$ (right)
estimate of $\protect\beta$ in the \textit{Fama} (1984) model; The null is $%
\protect\beta=1$ and the dashed ones are the 95\% confidence bands.}
\label{fig:rates}
\end{figure}

\begin{figure}[tbp]
\centering
\begin{tabular}{ccc}
\setlength{\itemsep}{-1.0cm} \psfig{file = 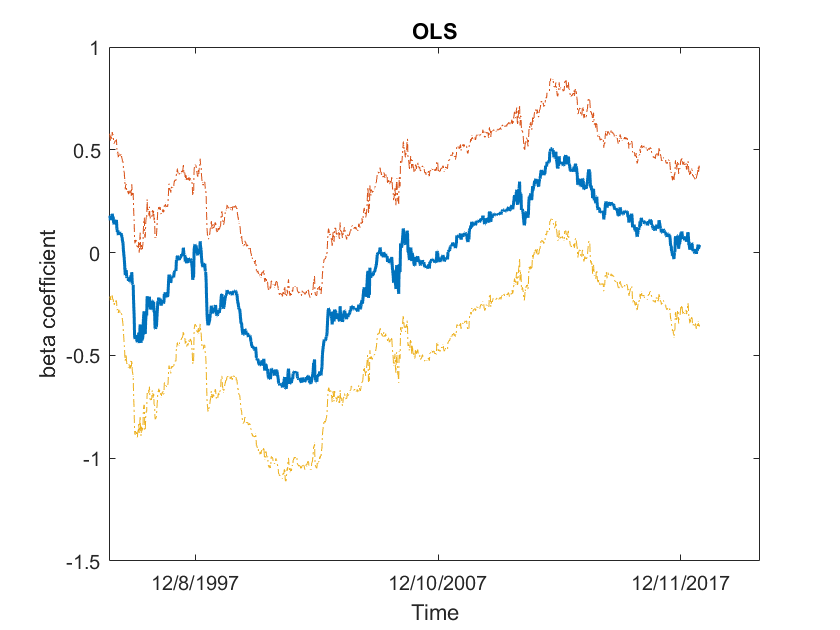, width = 7cm,
angle=0} & \psfig{file = 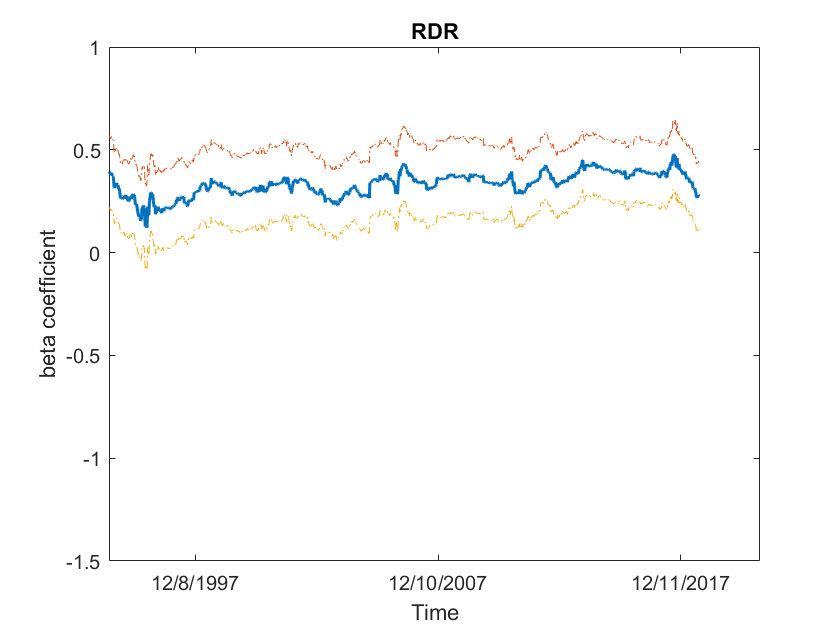, width = 7cm, angle=0} &  \\ 
New Zealand Dollar & New Zealand Dollar &  \\ 
\psfig{file = 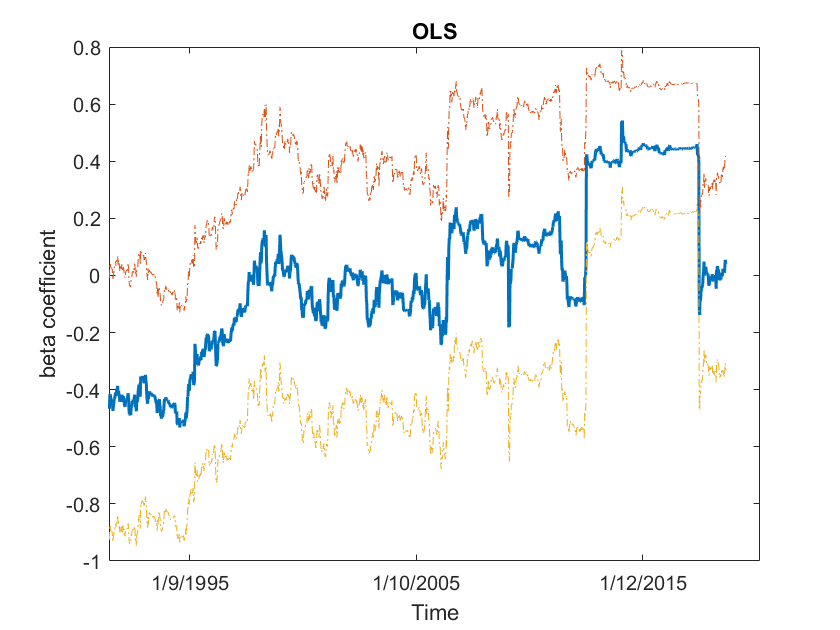, width = 7cm, angle=0} & \psfig{file =
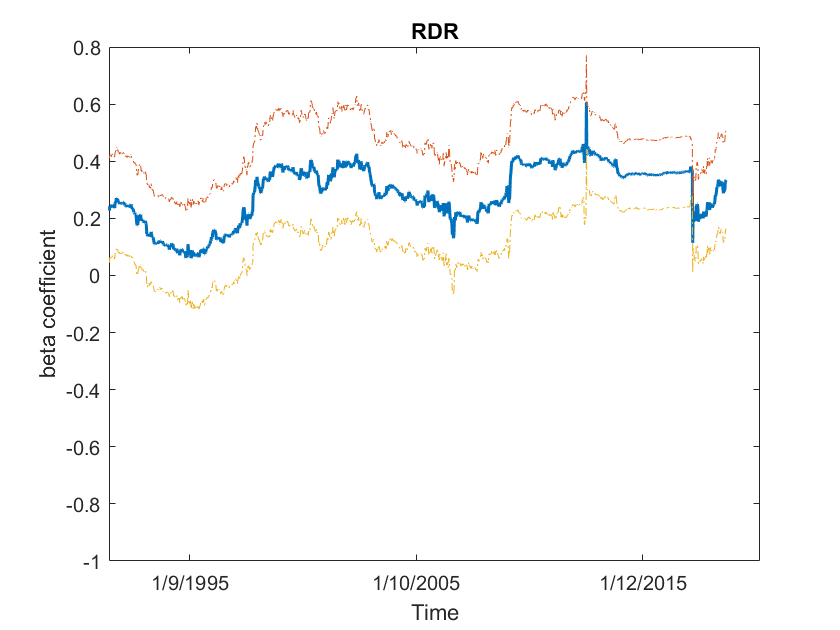, width = 7cm, angle=0} &  \\ 
Swiss Franc & Swiss Franc &  \\ 
\psfig{file = 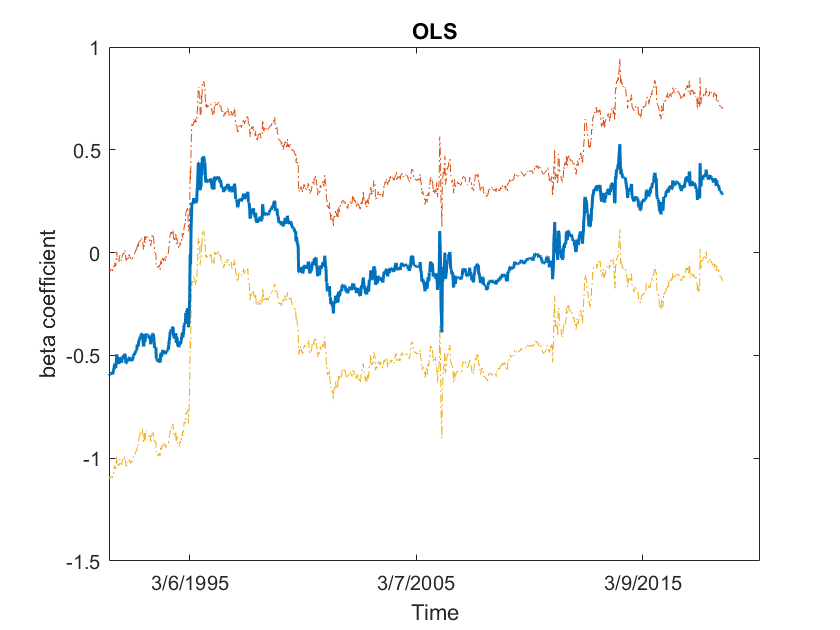, width = 7cm, angle=0} & \psfig{file =
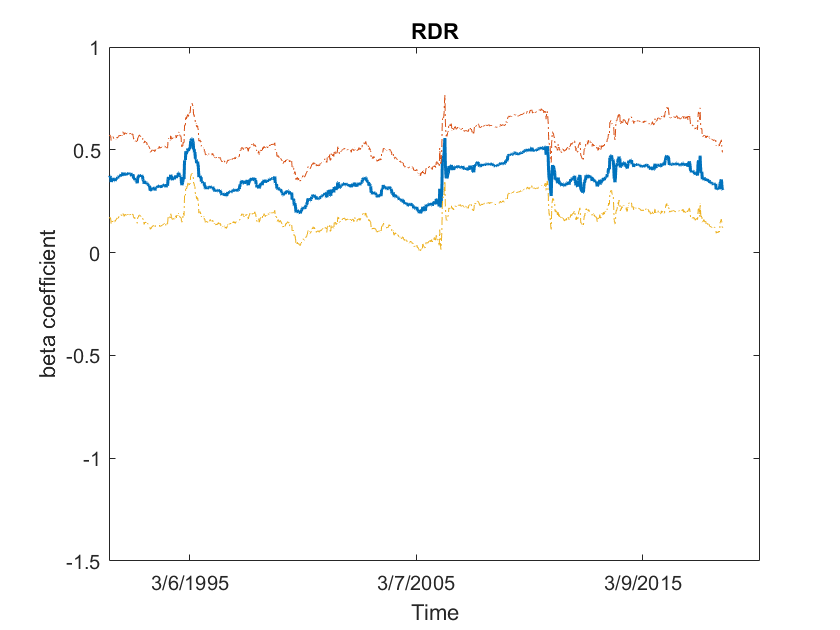, width = 7cm, angle=0} &  \\ 
UK Pound & UK Pound &  \\ 
&  & 
\end{tabular}
{\small {} }
\caption{The blue curve is the 5-year rolling OLS (left)/$RDynReg$ (right)
estimate of $\protect\beta$ in the \textit{Fama} (1984) model; The null is $%
\protect\beta=1$ and the dashed ones are the 95\% confidence bands.}
\label{fig:rates}
\end{figure}


\begin{thebibliography}{99}
\bibitem{Anderson1} Anderson, T., Bollerslev, T. 1995. Intraday Periodicity
and Volatility Persistence in Financial Markets, Journal of Empirical
Finance, 4, 115-158.

%\bibitem{Anderson1} Anderson, T., Bollerslev, T., Diebold, F.X., Labys, P.
%2003. Modeling and forecasting realized volatility. \textit{Econometrica} 
%\textbf{71:} 579-626.
%
%\bibitem{Anderson2} Anderson, T., Bollerslev, T., Diebold, F.X., Vega, C.
%2003. Effects of macro announcements: real time price discovery in foreign
%exhange. \textit{American Economic Review} \textbf{93:} 38-62.
$UIP$

\bibitem{Baillie4} Baillie R T. 1989. Econometric tests of rationality and
market efficiency. \textit{Econometric Reviews} \textbf{8:} 151-186.

\bibitem{Baillie2} Baillie R T, Bollerslev T. 1989. Common stochastic trends
in a system of exchange rates. \textit{Journal of Finance}, 44: 167-181.

\bibitem{Baillie1} Baillie R T, Diebold F X, Kapetanios G, Kim K H. 2022.
On robust inference in time series regression. \url{arXiv:2203.04080}, March, under revision.

\bibitem{Baillie1} Baillie R T, Lippens R E, McMahon, P C. 1983. Testing
rational expectations and efficiency in the foreign exchange market . 
\textit{Econometrica} \textbf{51:} 553-563.

\bibitem{Baillie5} Baillie R T, Osterberg W P. 1997. Central bank
intervention and risk in the forward premium. \textit{Journal of
International Economics. }\textbf{43:} 483-497.

\bibitem{Baillie6} Baillie R\ T, Kilic R, 2006. Do asymmetric and nonlinear
adjustments explain the forward premium anomaly? \textit{Journal of
International, Money and Finance.} \textbf{25:} 22-47.\ 

\bibitem{Bilson} Bilson J F O. 1981. The speculative efficiency hypothesis. 
\textit{Journal of Business. }\textbf{54:} 435-452.

\bibitem{Lustig} Burnside C, 2011. The cross section of foreign currency
risk premia and consumption growth risk:\ comment. \textit{American Economic
Review. }\textbf{101}: 3456-3476.

\bibitem{} Corbae D, Ouliaris S. 1988. Robust tests for unit roots in the
foreign exchange market. \textit{Economic Letters. }\textbf{22: }375-380.

\bibitem{Domowitz} Domowitz I, Hakkio C S. 1984. Conditional variance and
the risk premium in the foreign exchange market, \textit{Journal of
International Economics.} \textbf{19}: 47-66.

\bibitem{Fama.} Fama E F, 1984. Spot and forward exchange rates. \textit{%
Journal of Monetary Economics.} \textbf{19}: 319-338.

\bibitem{Frankel1} Frenkel J A, 1977. The forward exchange rate,
expectations and the demand for money: The German hyperinflation. \textit{%
American Economic Review}. \textbf{64}: 653-670.

\bibitem{Frenkel2} Frenkel J A, \ 1979. Further evidence on expectations and
the demand for money during the German hyperinflation. \textit{Journal of
Monetary Economics.} \textbf{5:} 81-96.

\bibitem{Frenkel3} Frenkel J A, Levich R M. 1975. Covered Interest
Arbitrage: Unexploited Profits?. \textit{Journal of Political Economy.} 
\textbf{83(2):} 325-338.

\bibitem{Grendander} Grenander U, 1981. \textit{Indirect Inference}.\ Wiley,
New York.

\bibitem{Hakkio} Hakkio C S, 1981. Expectations and the forward exchange
rate. \textit{International Economic Review} \textbf{22:} 663-678.

\bibitem{Hakkio} Hannan E J, Deistler M, 1988. \textit{The Statistical
Theory of Linear Systems}. SIAM..

\bibitem{Hansen} Hansen L P, Hodrick R J, 1980. Forward exchange rates as
optimal predictors of future spot rates: an econometric analysis. \textit{%
Journal of Political Economy,} \textbf{88}, 829--853.

\bibitem{HH} Hansen, L P, Hodrick, R J. 1983. Risk averse speculation in the
forward foreign exchange market: An econometric analysis of linear models. 
\textit{Exchange Rates and International Economics}, edited by Jacob A.
Frenkel.

\bibitem{Hodrick2} Hodrick R J. 1989. Risk, uncertainty and exchange rates 
\textit{Journal of Monetary Economics.} \textbf{23}: 433-459.

\bibitem{Ismael} Ismailov A, Rossi B, 2021. Uncertainty and deviations from
uncovered interest rate parity. \textit{Journal of International, Money and
Finance.} forthcoming.

\bibitem{Hodrick1} Kaminsky G, Peruga R, 1990. Can a time varying risk
premium explain excess returns in the forward markets for foreign exchange? 
\textit{Journal of International Economics}, \textbf{28}: 47-70.

\bibitem{levy} Levy E, Nobay A\ R, 1986. The speculative efficiency
hypothesis: A bivariate analysis. \textit{Economic Journal, Annual Supplement%
}, 109-121.1991.

\bibitem{Newey and West1} Newey W K, West K\ D, 1987. A simple positive semi
definite heteroscedasticity and autocorrelation consistent covariance
matrix. \textit{Econometrica} \textbf{55:} 703-708.

\bibitem{Newey, W.K. and K.D. West2} Newey W K, West K\ D, 1994. Automatic
lag selection in covariance matrix estimation, \textit{Review of Economic
Studies} \textbf{61:} 634--654.

\bibitem{Schwarz} Schwarz G E, 1978. Estimating the dimension of a model. 
\textit{Annals of Statistics} \textbf{6:} 641-464.

\bibitem{Taylor} Taylor M P, 1987. Covered interest parity: A high-frequency
high-quality data study. \textit{Economica} \textbf{54:} 429-438.
\end{thebibliography}
\end{document}